\documentclass[acmsmall]{acmart}
\AtBeginDocument{%
  }

\setcopyright{acmcopyright}
\copyrightyear{2024}
\acmYear{2024}
\acmDOI{XXXXXXX.XXXXXXX}

\acmPrice{15.00}
\acmISBN{978-1-4503-XXXX-X/18/06}

\usepackage{amsmath}

\usepackage{xspace}

\usepackage{multirow}

\usepackage{array}
\usepackage{enumitem}
\usepackage{graphicx}
\usepackage{subcaption}

\newcommand{\ATUA}{ATUA\xspace}
\newcommand{\APPR}{CALM\xspace}
\newcommand{\CALM}{CALM\xspace}
\usepackage[backgroundcolor=white,bordercolor=blue,linecolor=blue]{todonotes}

\newcommand{\RF}{$\mathcal{L}$\xspace}

\newcommand{\RFs}{$\mathcal{L}_{*}$\xspace}
\newcommand{\RFone}{$\mathcal{L}_{1}$\xspace}
\newcommand{\RFtwo}{$\mathcal{L}_{2}$\xspace}
\newcommand{\RFthree}{$\mathcal{L}_{3}$\xspace}
\newcommand{\RFfour}{$\mathcal{L}_{4}$\xspace}
\newcommand{\RFfive}{$\mathcal{L}_{5}$\xspace}

\newcommand{\CHANGED}[1]{\textcolor{black}{#1}}
\definecolor{byzantine}{rgb}{0.74, 0.2, 0.64}

\newcommand{\MAJORBEGIN}{}
\newcommand{\MAJOREND}{}

\newcommand{\MAJOR}[2]{#2}

\newcommand{\MAJORTWO}[2]{{#2}}

\begin{document}

\title{Testing Updated Apps by Adapting Learned Models}





\author{Chanh Duc Ngo}
\affiliation{%
  \institution{SnT Centre, University of Luxembourg}
  \streetaddress{JFK 29}
  \city{Luxembourg}
  \country{Luxembourg}}
\email{chanh-duc.ngo@uni.lu}

\author{Fabrizio Pastore}
\affiliation{%
  \institution{SnT Centre, University of Luxembourg}
  \streetaddress{JFK 29}
  \city{Luxembourg}
  \country{Luxembourg}}
\email{fabrizio.pastore@uni.lu}

\author{Lionel Briand}
\authornote{Part of this work was done while affiliated with University of Luxembourg.}
\affiliation{%
  \institution{Lero SFI Centre for Software Research, University of Limerick}
  \streetaddress{Tierney building}
  \city{Limerick}
  \country{Ireland}}
  \affiliation{%
  \institution{School of EECS, University of Ottawa}
  \city{Ottawa}
  \country{Canada}}
\email{lionel.briand@lero.ie}



\begin{abstract}
Although App updates are frequent and software engineers would like to verify updated features only, automated testing techniques verify entire Apps and are thus wasting resources. 

We present \emph{Continuous Adaptation of Learned Models (\APPR)}, an automated App testing approach that efficiently test App updates by adapting App models learned when automatically testing previous App versions. 
\APPR focuses on functional testing. Since functional correctness can be mainly verified through the visual inspection of App screens, \APPR minimizes the number of App screens to be visualized by software testers while maximizing the percentage of updated methods and instructions exercised. 

Our empirical evaluation shows that \APPR exercises a significantly higher proportion of updated methods and instructions than six state-of-the-art approaches, 
for the same maximum number of App screens to be visually inspected. Further, in common update scenarios, where only a small fraction of methods are updated, \APPR  is even quicker to outperform all competing approaches in a more significant way.

\end{abstract}




\maketitle


\section{Introduction}

Software applications for mobile devices (i.e., Apps) are updated frequently, mainly to improve the user experience and fulfill marketing strategies~\cite{Mcilroy:FrequentlyUpdatedApps:ESE:2016,Calciati:NewAppReseases:MSR:2018,Alvarez:WAMA:2019}. 
Our industry partners highlighted that in such scenario, where the time dedicated to development and testing is limited, it is important to focus testing effort on the features that have been modified and introduced in the new App version.

Unfortunately, automated App testing techniques do not target updated features but exercise whole Apps and cover their implementation only partially (e.g., they exercise around half of the App methods~\cite{Wang:EmpStudy:2018,Choudhary-AutomatedTestInputGeneration-ASE-2015}). 
When coverage is limited, regression test selection techniques~\cite{Sharma:QADroid:ISSTA:2019,Choi:DetReduce:ICSE:2018} are unlikely to help engineers in selecting test cases that exercise the updated features. Therefore, the automated testing of updated features remains an open problem.
Further, existing techniques detect only crashes or data loss~\cite{RiganelliMRMM20} though 
a recent study on functional faults affecting Android Apps reports that 95\% of the failures likely require visual inspection to be detected.
Among these, content related issues account for 21\%, structure related issues 40\%, incorrect interaction 19\%, and functionality not taking effect 12\%) ~\cite{FuntionalBugsAndroid}.
Unfortunately, visual inspection of App outputs is practically infeasible when automated testing tools generate a large number of test inputs, each one leading to a new output screen to be inspected.

Our previous work has shown that static and dynamic program analyses drive model-based App testing towards maximizing the coverage of updated methods while using a limited number of test inputs~\cite{ATUA}. We named our previous approach ATUA; for a same number of App screens to be exercised, it outperforms state-of-the-art (SOTA) approaches 
in terms of code coverage. 

Although ATUA demonstrated to be more effective than 
approaches not focused on App updates, it does not reuse App models across versions, which makes the test process inefficient (e.g., for every App version, it may resort to random exploration to trigger Window transitions not identified by static analysis).
In the literature,
inferred models have been reused to repair test scripts~\cite{SITAR}, execute test cases on different platforms~\cite{MAPIT,ReRc}, and automate regression testing~\cite{7381848}. 
Unfortunately, the only approach reusing models across versions is Fastbot2~\cite{Fastbot2}, a recent approach that reuses a probabilistic model 
learned in a previous version to drive testing in a newer version. However, our empirical results (see Section~\ref{sec:empirical}) show that, since it does not integrate static analysis, it cannot effectively target updated features.

%

We present \emph{Continuous Adaptation of Learned Models (\APPR)}, an 
App testing technique that efficiently tests updated Apps by relying on models learned with previous App versions. 
\APPR leverages ATUA to select test inputs that exercise updated methods.
However, \APPR improves over ATUA by combining dynamic and static program analysis to adapt and improve 
the model learned when testing a previous App version. 
The reuse of an existing model enables \APPR to efficiently use the test budget to exercise updated methods rather than to determine, with random exploration, how to reach Windows already reached in previous App versions. \MAJOR{1.1}{Like ATUA, \APPR aims at detecting functional faults leading to erroneous App outputs but is not targeting crashes. Since crashes can be automatically detected, existing approaches (e.g., Monkey) can readily be used for that purpose. \APPR, instead, optimizes the test budget not only to quickly reach the App states enabling the execution of updated methods but also to minimize the number of outputs to be inspected by engineers to detect  functional faults.}

Before testing a new App version, \APPR relies on static program analysis to identify changes in the App GUI that should be reflected in the App model. This includes, for example, removing  state transitions triggered by Widgets no longer present in the updated App. 
In addition, it integrates heuristics for the runtime adaptation of App models to make model reuse effective. Precisely,
it introduces \emph{layout-guarded abstract transitions} to deal with non-determinism; it derives \emph{probabilistic Action sequences}
to deal with \emph{state explosion}; it detects model states that are new but compatible with previously executed Action sequences (i.e., \emph{backward-equivalent}); it relies on \emph{online and offline model refinement} to identify and remove obsolescent model states.
Finally, \APPR identifies and provides engineers 
with only the output screens rendered by the App after an updated method had been exercised, thus greatly minimizing test oracle costs.

Our empirical evaluation shows that, for a one-hour test budget, \CALM exercises a significantly larger percentage of updated methods and instructions than SOTA tools (ATUA, 
Monkey~\cite{monkey},  APE~\cite{Gu:APE:ICSE:2019}, Fastbot2, TimeMachine~\cite{TimeMachine}, and Humanoid~\cite{Humanoid}).
For a same maximum number of screen outputs to be visualized, \CALM outperforms the second-best SOTA approach by 6 percentage points (pp). Most importantly, this difference keeps increasing with the test budget and is even larger (13 pp) for quick test sessions with updates of small size, which are by far the most frequent.

Section~\ref{sec:background} introduces background technologies.
Section~\ref{sec:approach} describes the proposed approach.  
Section~\ref{sec:empirical} reports on the results of our empirical evaluation.  
Section~\ref{sec:related} discusses related work.
Section~\ref{sec:conclusion} concludes the paper.

\section{Background}
\label{sec:background}

\subsection{Model-based App Testing with \ATUA}
\label{back:atua}

Most of the App testing automation approaches reported in the literature are \emph{model-based} (i.e., they infer an App model that is used to drive testing~\cite{Amalfitano:AndroidTestingSurvey:SQJ:2018}).
They differ for the type of analysis used to identify states and transitions (i.e., dynamic~\cite{stoat17,Borges-Droidmate2-ASE-2018} or static~\cite{yang-jase18}), the abstraction functions used (i.e., predefined~\cite{stoat17,Borges-Droidmate2-ASE-2018} or  adaptable~\cite{Gu:APE:ICSE:2019}), and their model exploration strategy (i.e., offline~\cite{stoat17} or online~\cite{Borges-Droidmate2-ASE-2018}). 

We rely on ATUA~\cite{ATUA}, a recent model-based solution that integrates multiple analysis strategies. To focus on updated (modified and new) methods, it combines static analysis (to determine the inputs that execute updated features) and random exploration (to overcome the limitations of static analysis). To generate a reduced set of test inputs, it relies on dynamically-refined state abstraction functions (to determine when distinct inputs lead to a same program state) and information retrieval techniques (to identify dependencies among App features).

\begin{figure*}[tb]
  \centering
	\includegraphics[width=15cm]{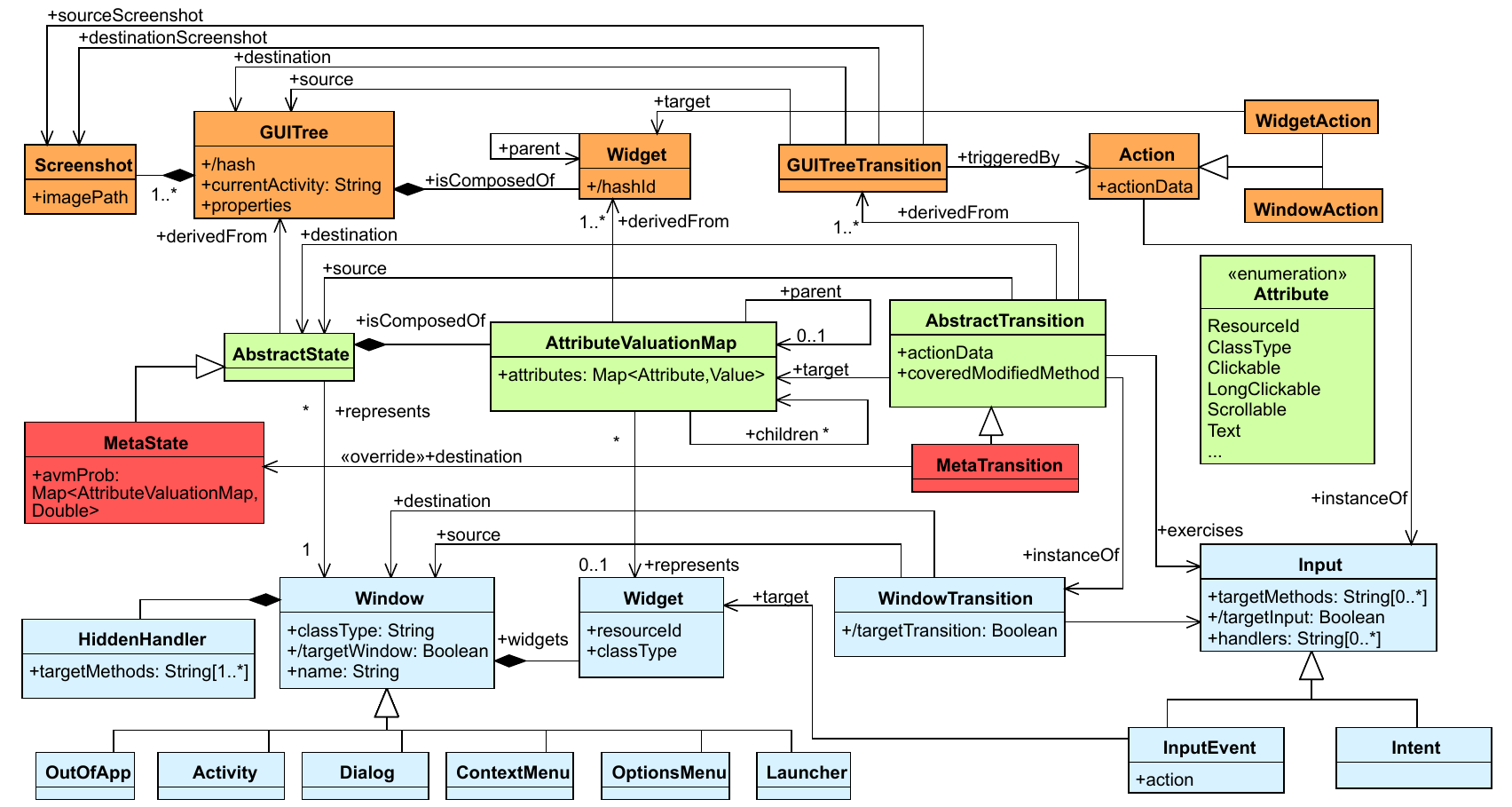}
      \caption{App Model Metamodel. Colors are used to group classes belonging to a specific metamodel component: GSTG (orange, top), DSTG (green, middle), EWTG (light blue, bottom). Classes in red are specific to \APPR.}
      \label{fig:app:model:metamodel:FULL}
\end{figure*}

The testing activity performed by \ATUA is driven by an App model whose  metamodel is shown in Figure~\ref{fig:app:model:metamodel:FULL}. It consists of three parts: (1) an Extended Window Transition Graph (EWTG), (2) a Dynamic State Transition Graph (DSTG), and (3) a GUI State Transition Graph (GSTG). 
The \emph{EWTG} is extracted by relying on the static analysis tool Gator~\cite{yang-jase18}; it models the sequences of Windows being visualized after 
triggering specific Inputs (Events or Intents). 
The EWTG extends Gator's WTG with, for every Input, the list of target methods that may be invoked during the execution of the input handler.
The \emph{GSTG} captures the \emph{GUITree} (i.e., the 
hierarchy tree of widgets and their properties) 
that might be visualized after an Action is performed on the GUI. 
An Action is an instance of an Input (e.g., click on a specific Button widget). 
Finally, the \emph{DSTG} models the \emph{AbstractStates} of the visualized Windows and the \emph{AbstractStateTransitions} triggered by events. 
\CHANGED{AbstractStates are identified by a state abstraction function. 
The DSTG helps determine a valid and minimal sequence of Actions necessary to reach a specific Window.} 

The App model is used to drive testing with the objective of exercising a set of \emph{target methods} (i.e., the methods introduced or modified in the updated App version).
To test an App, \ATUA identifies the sequence of Actions necessary to reach a target Window, that is, a Window in which some Actions (hereafter, \emph{target Actions}) may trigger the execution of target methods.
Precisely, \ATUA  identifies the shortest Action sequence by relying on a breadth-first traversal that considers both WindowTransitions (in the EWTG) and AbstractTransitions (in the DSTG). 
AbstractTransitions are prioritized because more accurate. Indeed, a certain WindowTransition may be enabled only if the App is in a specific state; instead, AbstractTransitions indicate that a certain Window can be reached from a certain AbstractState, based on previous testing results.
Once in a target Window, \ATUA triggers target Actions.

During testing, \ATUA identifies AbstractStates through \emph{abstraction functions} that are automatically selected for each Window of the App under test.
To this end, \ATUA relies on a predefined set of \emph{reducers} (see Table~\ref{table:reducers}), i.e., functions that
extract the value of a widget property~\cite{Gu:APE:ICSE:2019}.
For each Widget, \ATUA keeps track of the applied reducers and their outputs 
in a map, called AttributeValuationMap (AVM).
The AVM has a cardinality attribute indicating how many Widgets have the same attribute valuations. 
Two AbstractStates differ when at least one value differs across their respective AVMs.


\begin{table}[tb]
\caption{\ATUA reducers and abstraction functions. \RFfour and \RFfive match \RFtwo but they apply \RFone and \RFtwo reducers to Widgets' children, respectively.}
\label{table:reducers}
\footnotesize
\begin{tabular}{
|@{\hspace{1pt}}>{\raggedleft\arraybackslash}p{10mm}@{\hspace{1pt}}|
@{\hspace{1pt}}>{\arraybackslash}p{60mm}@{\hspace{1pt}}|
@{\hspace{1pt}}>{\raggedleft\arraybackslash}p{14mm}@{\hspace{1pt}}|
}
\hline

\textbf{Reducer}&\textbf{Property extracted}&\textbf{Function}\\
\hline
 $R_{RID}$&Resource ID.&\RFone,\RFtwo,\RFthree\\

$R_{CN}$&Class name.&\RFone,\RFtwo,\RFthree\\

$R_{CD}$&Value of \emph{Content description}.&\RFone,\RFtwo,\RFthree\\

 $R_{P}$&Value of \emph{Password}.&\RFone,\RFtwo,\RFthree\\

 $R_{C}$&Value of \emph{Clickable}.&\RFone,\RFtwo,\RFthree\\

$R_{LC}$&Value of \emph{Long Clickable}.&\RFone,\RFtwo,\RFthree\\

$R_{S}$&Value of \emph{Scrollable}.&\RFone,\RFtwo,\RFthree\\

$R_{Ch}$&Value of \emph{Checked}.&\RFone,\RFtwo,\RFthree\\

$R_{E}$&Value of \emph{Enabled}.&\RFone,\RFtwo,\RFthree\\

$R_{S}$&Value of \emph{Selected}.&\RFone,\RFtwo,\RFthree\\

$R_{I}$&True if it is an input field.&\RFone,\RFtwo,\RFthree\\

$R_{T}$&Value of \emph{Text}.&\RFtwo,\RFthree\\

$R_{HC}$&True if the widget contains one or more children.&\RFthree\\

\hline
\end{tabular}
\end{table}%

To minimize non-determinism in the DSTG,
during testing, when \ATUA observes that a same Action may bring the App into two distinct AbstractStates, 
it refines the abstraction function for the Window in which the Action had been triggered.
Refinement is performed by increasing the number of reducers used to extract the information captured in AVMs. It relies on five different set of reducers for the state abstraction function; they are named as \RFone to \RFfive and described in Table~\ref{table:reducers}.

In \ATUA, testing is performed in three phases, the first aims at maximizing the number of target transitions being exercised (i.e., it triggers every Input once), the second aims at increasing testing of the less exercised target transitions (i.e., it triggers again Inputs that may reach target methods not fully covered), the third aims at exercising target Windows that depend on specific App states reached in related Windows (i.e., it exercises a target Window just after a related Window). 

\ATUA uses part of the test budget to overcome the limitations of static analysis through dynamic analysis. Indeed, \ATUA resorts to random exploration when WindowTransitions are infeasible (i.e., when it is necessary to find the AbstractState in which they are enabled) and when Windows are unreachable (e.g., if static analysis does not detect any WindowTransition reaching a specific Window). Also, \ATUA often refines abstraction functions to eliminate non-determinism. In summary, the AbstractStates and AbstractTransitions identified during testing complement the information captured in the EWTG. 
\MAJOR{2.1}{Note that our previous empirical results~\cite{ATUA} have shown that the integration of static and dynamic analysis implemented by ATUA leads to higher test effectiveness than dynamic analysis alone (i.e., what implemented by DM2~\cite{Borges-Droidmate2-ASE-2018}, the backbone of ATUA).}
Unfortunately, by creating App models from scratch for every App version, \ATUA may perform the same model refinements when testing each App version thus wasting test budget that could be used to increase code coverage.


\subsection{RCVDiff}
\label{rcvdiff}


\MAJOR{R3.2}{To reuse App models across versions, CALM must identify what App GUI components were left unchanged. To this end, it identifies the differences in the EWTGs of two App versions by relying on RCVDiff, which }
is a toolset for the identification of model differences~\cite{Protic2011,VandenBrand2010}. 
RCVDiff processes two RCVModels;
an RCVModel is a collection of \emph{MElements}, which contain \emph{MAttributes} and \emph{MReferences}. Mapping a UML object diagram to an RCVModel is straightforward.





RCVDiff relies on a tree-matching algorithm that is applied to a tree data structure derived from the RCVModel. The transformation is enabled by the tree-like structure of the RCVModel, following sub-elements relations. 
The matching algorithm works in three steps.
First, RCVDiff proceeds bottom-up to identify, for each element in the first model, a list of corresponding elements in the second model. 
The lists is obtained by identifying, for each element, a set of elements with matching attributes, references, and sub-elements. To identify matching  attributes, RCVDiff relies on a set of predefined similarity functions and thresholds.
References match when the referenced elements' attributes match.
In the second step, RCVDiff proceeds top-down with the objective of maximizing the number of matching pairs.
Finally, it relies on a bottom-up pass to identify changed, added, and deleted Elements.

\section{Proposed Approach: \APPR}
\label{sec:approach}

\begin{figure}[tb]
\includegraphics[width=8.4cm]{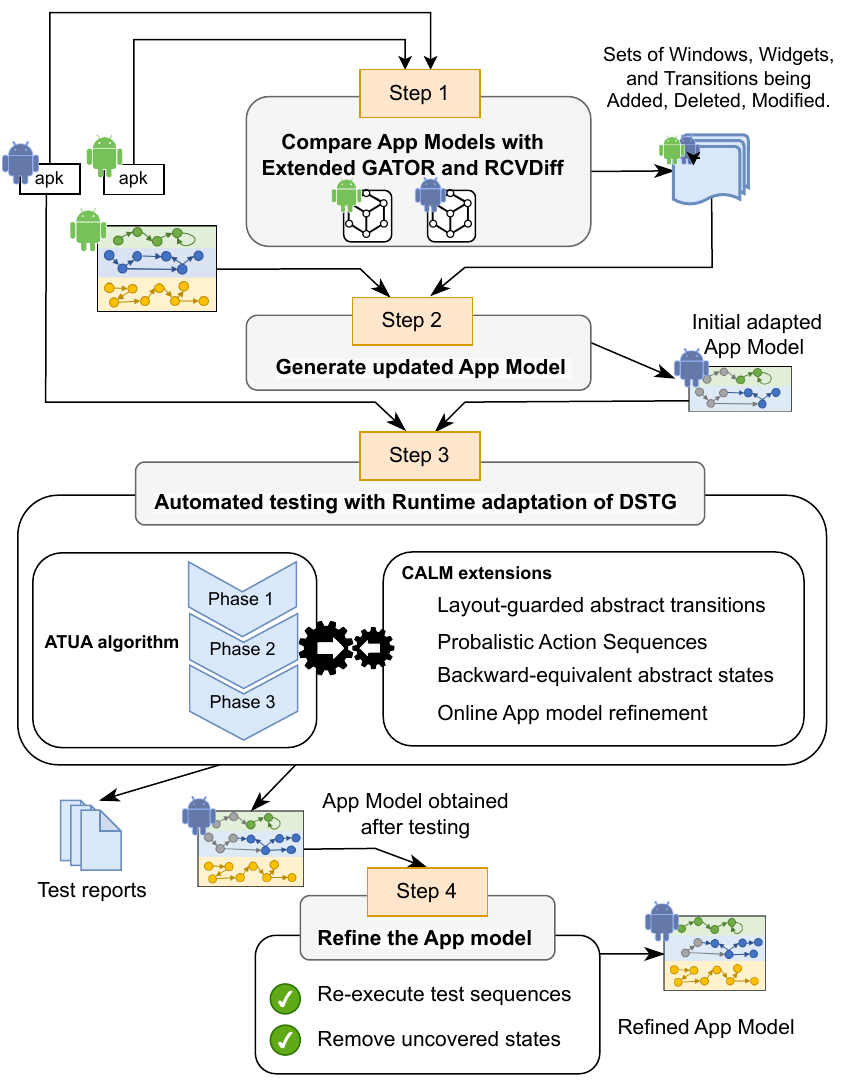}

\caption{\APPR App testing process}
\label{fig:updateprocess}
\end{figure}

\APPR 
supports engineers in 
testing updated Apps by relying on App models that are incrementally constructed and adapted, version after version. Similar to \ATUA, 
\APPR aims to exercise all  the \emph{target methods} (see Section~\ref{sec:background}) of an App version.
For the first version of the App under test, 
\APPR treats each method as a target method, and starts from an empty App model.
For every App version following the first, \APPR relies on the App model produced for the previous App version (\emph{base App}).

\MAJOR{R3.8}{Since App updates may change or remove the  GUI Windows and Widgets being visualized, thus changing how one interacts with different versions of an App, the App model should not include any information that is obsolete, not to undermine test effectiveness. For example, it should not be possible to select Action sequences that traverse Windows that are no longer present in the updated App version. In practice, the App model derived from a previous App version cannot be reused as-is. However, \CALM should maximize the amount of information that is preserved from previous App models; specifically, since App updates may include refactorings (e.g., renaming Windows and moving Widgets across Windows), instead of simply identifying what App model elements are missing in the updated App, \CALM should determine what App model elements match across two App versions.} 

\MAJOR{R3.8}{To decide what elements of the App model should be reused across App versions, we should also take into account how testing is driven in \ATUA, which is the backbone of \CALM. In \ATUA, the selection of Action sequences is based on the EWTG, which provides information about what Inputs of a Window may trigger a target method, and the DSTG, which indicates in which AbstractStates a certain Input is enabled (e.g., a button is visualized). The GSTG, instead, is used to keep track of the App screens visualized during testing when manual investigation is needed. Also, the GUITrees in the GSTG are used to refine AbstractStates (i.e., to split one AbstractState using a refined \RF). 
Since the GSTG includes cosmetic information (e.g., the position of a Widget), it is very likely to change across versions. Further,  GSTG elements are used only to refine abstract states, which implies that reusing GSTG elements from previous versions may lead to AbstractStates that no longer exist. Therefore, \CALM discards the GSTG of the inherited App model. The EWTG and DSTG should instead be kept because they capture high-level information that is likely to remain unchanged across versions.
Since any change in the EWTG is reflected on the DSTG (e.g., we cannot have AbstractStates for a Window no longer present in the App), \CALM needs to determine what elements belonging to the EWTGs of the previous and updated App versions match, in order to determine what information (i.e., Transitions, AbstractStates, and AbstractTransitions) should be preserved in the DSTG. Further, since some changes can be determined only at runtime (i.e., what elements are visualized), \CALM further refines the App model during testing, as we outline next.}

\APPR works in four steps, shown in Figure~\ref{fig:updateprocess}. In Step 1, \APPR relies on an extension of RCVDiff to 
compare the EWTGs generated by \ATUA's extended Gator for the base  and updated App.

In Step 2, \APPR relies on the identified differences to generate an updated App model by adapting the EWTG and the DSTG of the base App model. 
By reusing the DSTG generated for the base App, \APPR aims to optimize testing efficiency by minimizing the effort spent to generate AbstractStates and AbstractTransitions for the updated App. 

In Step 3, \APPR relies on an extended version of the \ATUA testing process to test the updated App. CALM's extensions maximize the effectiveness of model reuse by enabling the execution of Action sequences derived from previous App models, even when the AbstractStates observed in the two versions present differences.
Further, \APPR updates the App model to reflect the actual behaviour of the App under test. 
\MAJORTWO{2.1b}{CALM inherits from ATUA the capability of relying on random exploration to overcome some limitations of static analysis; specifically, it relies on random exploration when it cannot reach a desired target Widget, which could happen for Windows introduced by the updated App or when the behaviour of the updated App changed. However, since CALM leverages the DSTG generated for the base App, which includes the information (i.e., AbstractStates and AbstractTransitions) collected from previous random explorations, the test budget spent to perform random exploration with CALM is lower than with ATUA.}

\begin{figure}
\hspace{7mm}\includegraphics[width=7.4cm]{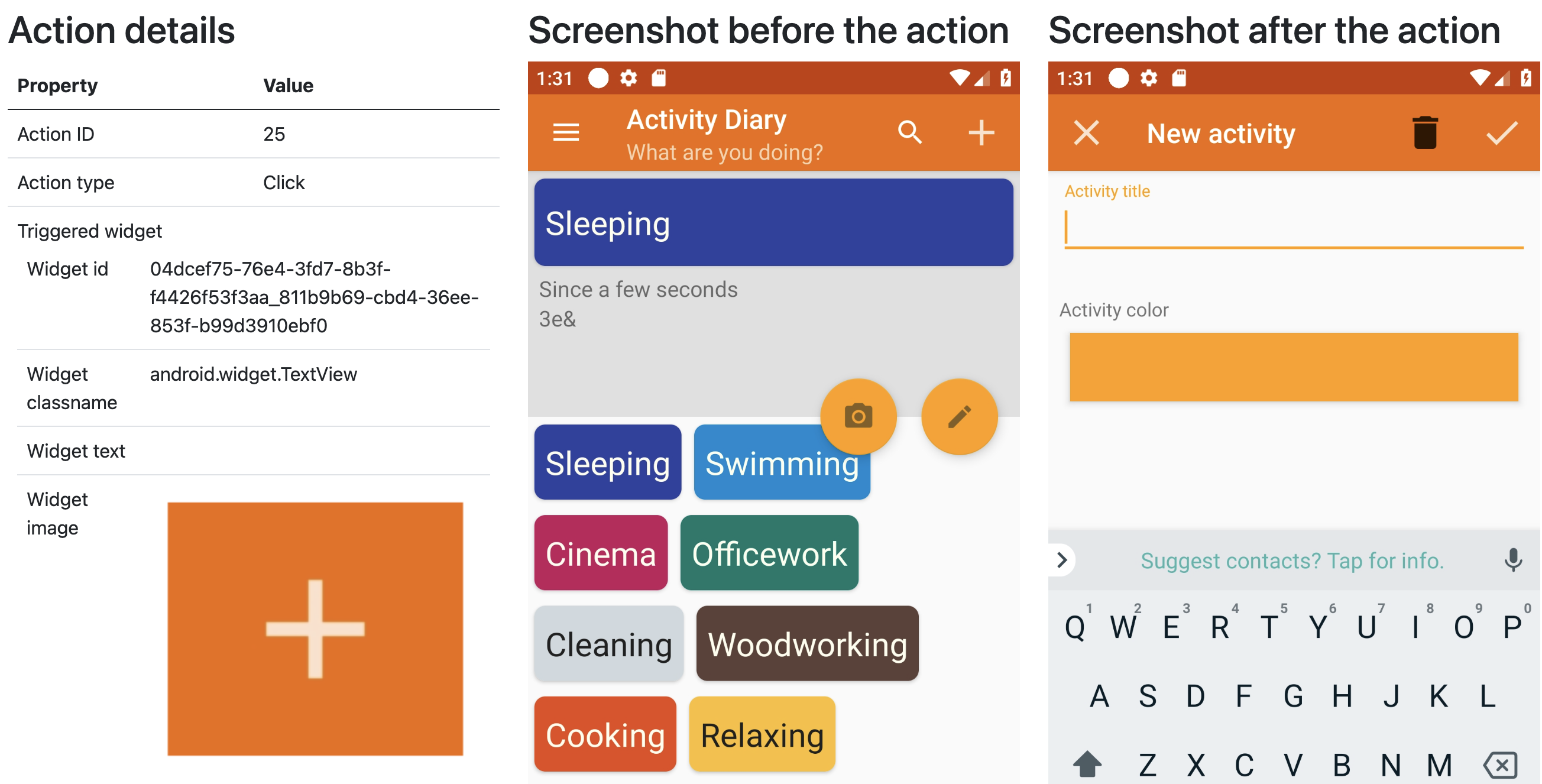}
\caption{Example of action output provided to the end-user}
\label{fig:actionOutput}
\end{figure}

The output of Step 3 is an App model for the updated App version. 
\APPR generates a report with a set of triples $<$GUI screenshot, target action, GUI screenshot$>$ reporting for every target Action (see Section~\ref{sec:background} for definition) triggered by \APPR the screenshot before and after the execution of the action. An example is shown in Figure~\ref{fig:actionOutput}. To avoid wasting engineers' time, only actions that increase code coverage 
are reported; we refer to such actions as \emph{Unique Target Actions} (\emph{UTAs}). The generated triples support crowdsourcing-based oracles (e.g., they can be shared among a set of App users to determine if the output is functionally correct or not~\cite{Pastore:CrowdOracles}). Further, engineers can also visualize, from the GSTG,  the sequence of inputs and outputs terminating with the triple shown in the report. Determining the best strategy to support end-users in fault detection is future work.

In Step 4, after testing, \APPR refines the App model to eliminate infeasible paths due to AbstractStates that are unreachable. Those are either AbstractStates from the base App model not observed when testing the updated App or AbstractStates introduced when testing the updated App but becoming quickly obsolete.
The  output of Step 4 is a refined App model to be used when testing the next App version.

In the following Sections, 
before detailing the \APPR steps, we first discuss how model reuse can exacerbate state abstraction's limitations and reduce testing effectiveness.

\subsection{Step 1: Detect EWTG differences}



\MAJOR{R3.8}{To determine what App model elements match, \APPR compares the EWTGs of the previous and updated App versions by relying on RCVDiff.} To this end, it generates an RCVModel instance that captures the EWTG elements.
To identify differences, we extended the RCVDiff algorithm as follows.
First, our RCVDiff extension looks for elements (Window, Widget, Transition) that present matching attributes and references. 
Second, to determine what elements of the base App model had been replaced in the updated App model, our RCVDiff extension looks for additional elements that may \emph{correspond} (e.g., because of a class renaming) by relying on the following:
\begin{itemize}[leftmargin=*]
    \item Since a Widget could be moved to another container (i.e., its parent changed), two Widgets correspond when all their attributes, except \emph{parent}, match. Also, since a Widget could be replaced with another one implementing similar features (e.g., a button replaced by a clickable image), they correspond when all their attributes, except \emph{className}, match.
    \item To correspond, two Windows should extend the same Window Type 
    and other properties should match (e.g., class). 
    \item To correspond, two Transitions should start from matching sources (i.e., Windows) and trigger the same action on a matching Widget (e.g., a Button). When the  destination does not match, the updated transition simply reflects a change in the App behaviour.
        
    \item To correspond, element attributes
        should have high string similarity; precisely, we rely on the Levenshtein ratio with a threshold of 40\%. 
        The chosen threshold has been empirically demonstrated to be appropriate~\cite{Pan2020a}, as it enables handling cosmetic changes (e.g., fixing typos in labels).
        For XPaths, which capture the position of a Widget in the containing Window, 
        we use a token-based distance: we split each string into tokens (separator is \emph{`/'}) and compute the cosine similarity distance~\cite{Yu2016} between the two token sets.
     \end{itemize}
     

\begin{figure}
\includegraphics[width=14cm]{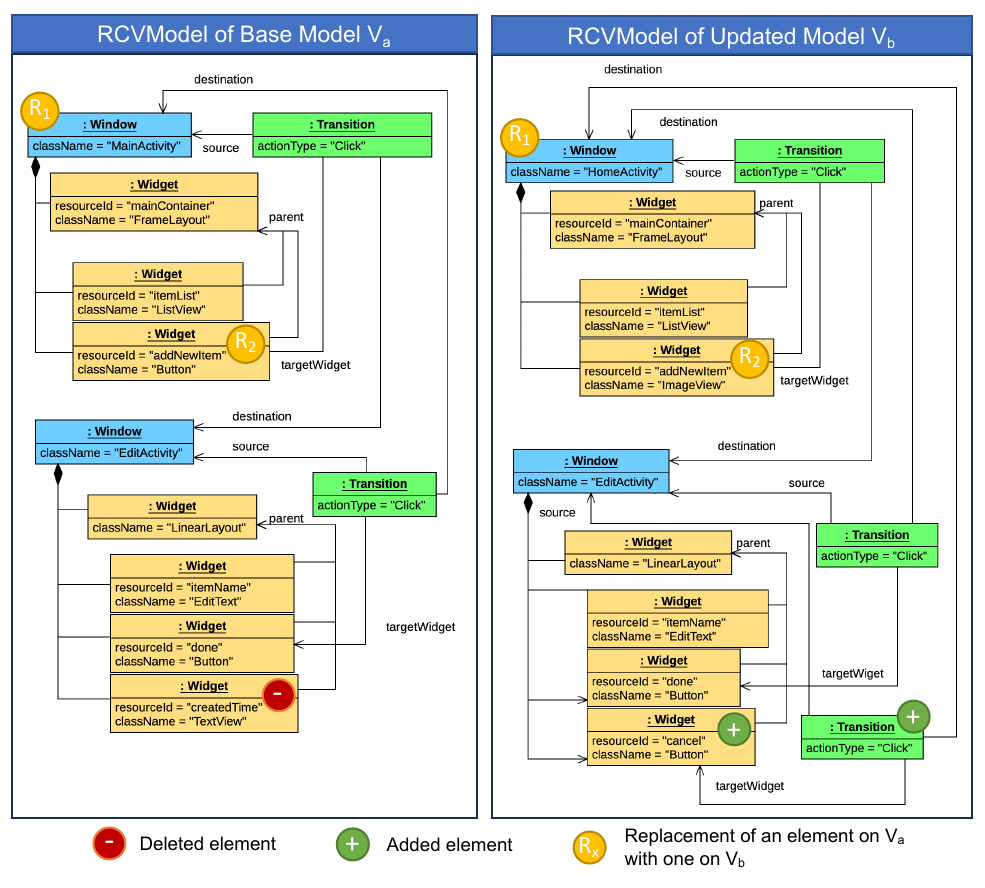}

\begin{minipage}{14cm}
\begin{footnotesize}
\textbf{Note:}  Our RCVDDiff extension reports that the Windows named \emph{MainActivity} in $V_a$ and \emph{HomeActivity} in $V_b$ correspond because they have the same type and their other properties match (they are not listed in the Figure); it indicates that the Window has been renamed.
\CALM thus correctly records that the Window \emph{MainActivity} in $V_a$ has been replaced by Window \emph{HomeActivity} in $V_b$.
Further, our RCVDDiff extension reports that the Widget named \emph{addNewItem} in $V_a$ corresponds to the homonymous widget in $V_b$ because all their attributes except their class name match; since the \emph{className} attribute of the widget \emph{addNewItem} has been changed from \emph{Button} to \emph{ImageView}, it indicates that a button Widget has been replaced by an image. 
\CALM thus records that the Widget \emph{addNewItem} in $V_a$ has been replaced by the Widget \emph{addNewItem} in $V_b$.
Finally, RCVDDiff detects that, in the \emph{EditActivity} Window, a Widget has been deleted (i.e., the TextView named \emph{createdTime}), while a Widget (i.e., the Button named \emph{cancel}) and a transition triggered by it have been added.
\end{footnotesize}
\end{minipage}
\caption{An example of RCVDiff Model of EWTGs belonging to two App versions.}
\label{fig:rcvdiff:wtgdiff}
\end{figure}

\APPR processes the RCVDiff output to identify Windows, Widgets, and Transitions being added, removed, or replaced; replaced elements are elements of the base App model with a corresponding element in the updated App model. Figure~\ref{fig:rcvdiff:wtgdiff} shows an example output.

\MAJORTWO{2.1a}{When comparing the EWTGs of the base and updated versions, \CALM ignores elements (e.g., Widgets, Windows, and WindowTransitions) that remain undetected by Gator but are observed at runtime thanks to dynamic analysis and random exploration. Such solution prevents CALM from considering such elements as deleted simply because they are not identified with static analysis. A common case is that of Dialogs; indeed, unlike Activities, which are declared explicitly in the Android manifest file, Dialogs are triggered in different ways. For example, if a dialog is created by an imported library, Gator cannot identify it.}

\subsection{Step 2: Generate an Updated App model}
\begin{figure}
    \centering
    \includegraphics[width=14cm]{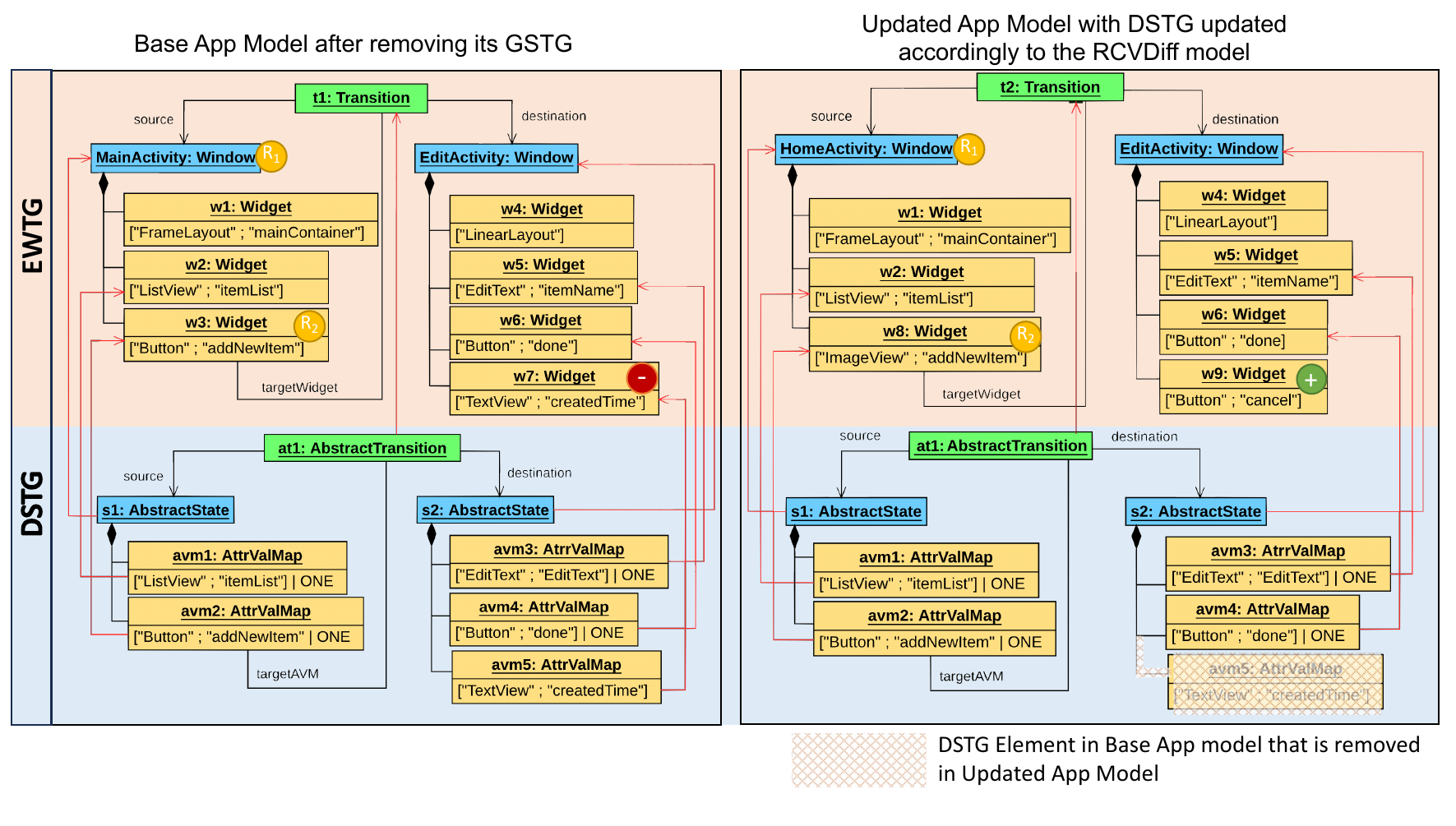}
    \caption{Illustration of how DSTG of Base App model is adapted in Updated App model accordingly to the RCVDiff model in Figure~\ref{fig:rcvdiff:wtgdiff} }
    \label{fig:step2-example}
\end{figure}
\APPR performs four tasks to create the updated App model to be used when testing the updated App: (1) copies the base App model, (2) removes all the GSTG elements, (3) replaces the base EWTG with the updated EWTG, and (4) updates the DSTG. 
Since the first three activities are straightforward, below, we describe how \CALM updates the DSTG.



\APPR removes from the DSTG all the items associated to the elements deleted from the EWTG, which are: all the AbstractStates associated with deleted Windows, all the AVMs associated with deleted Widgets, all the AbstractTransitions associated with deleted Transitions. Further, it removes all the elements that become disconnected from the rest of the DSTG.

Window replacements are caused by Windows renaming; therefore, \APPR assigns the replaced Window's AbstractStates to the replacing Window. Similarly, we assign replaced Widgets' AVMs to replacing Widgets. For Transition replacements, since they indicate a change in the source or destination Window but there is no mean to determine the mapping to an AbstractState for that Window, we simply remove the AbstractTransitions associated to the replaced Transition.

Added elements do not lead to any update of the DSTG because they were not present in the  base App. 

Figure~\ref{fig:step2-example} demonstrates how \CALM integrates the DSTG of a base App model into an updated App model by relying on the information provided by the RCVDiff model (i.e., the one in Figure~\ref{fig:rcvdiff:wtgdiff}, in this example). 
In the updated App model, the AttributeValuationMap \emph{createdTime} (i.e., \emph{avm5}) is removed from the AbstractState $s_{2}$ because, in the base App model, avm5 was associated with the Widget $w_{7}$, which has been removed from the updated App. 
In the updated App model, the AbstractState $s_1$ is reassigned to the \emph{HomeActivity} Window  because the \emph{HomeActivity} Window replaces the \emph{MainActivity} Window of the base App model. 
Still in the updated App model, the AttributeValuationMap \emph{addNewItem} (i.e., \emph{avm2}) is reassigned to the widget $w_8$, which has type \emph{ImageView}; such change depends on $w_8$  being a replacement for Widget $w_3$, which is of type \emph{Button}. Please note that preserving AVMs enable \CALM to preserve the AbstractTransitions $at_1$, which brings the App to the AbstractState $s2$ from $s1$ when clicking on the \emph{addNewItem} widget. 
Finally, since Widget $w_9$ was added to the updated App model, it has no DSTG element assigned to it. 
The other DSTG elements in the updated App model remain associated to the same elements in the base App model; or, more precisely, they are associated to EWTG elements of the updated App model that match the ones in the base App Model. 

\subsection{Step 3: Automated testing with runtime DSTG adaptation}

\MAJOR{2.6,3.4}{During testing, \CALM, like \ATUA, derives a sequence of Actions to be triggered to reach a target Window (in Phase 1) or a target AbstractState (in Phases 2 and 3).
Such sequence also specifies the AbstractState that is expected after every Action; if this expected state is not reached (e.g., because of non-determinism), the rest of the sequence is not executed. Indeed, it makes no sense to execute Actions whose  preconditions (e.g., a visible Widget) do not hold. 
When such state mismatch is observed, \APPR derives a new Action sequence that reaches the target Window/AbstractState from the current AbstractState. Unfortunately, when models are reused, state abstraction mechanisms often lead to such state mismatches. 
In this Section, we describe the solutions integrated into the \CALM testing step to prevent such state mismatches.}


\subsubsection{Layout-guarded abstract transitions mitigating non-deterministic AbstractTransitions}{\ }

\begin{figure}[tb]
         \includegraphics[width=13cm]{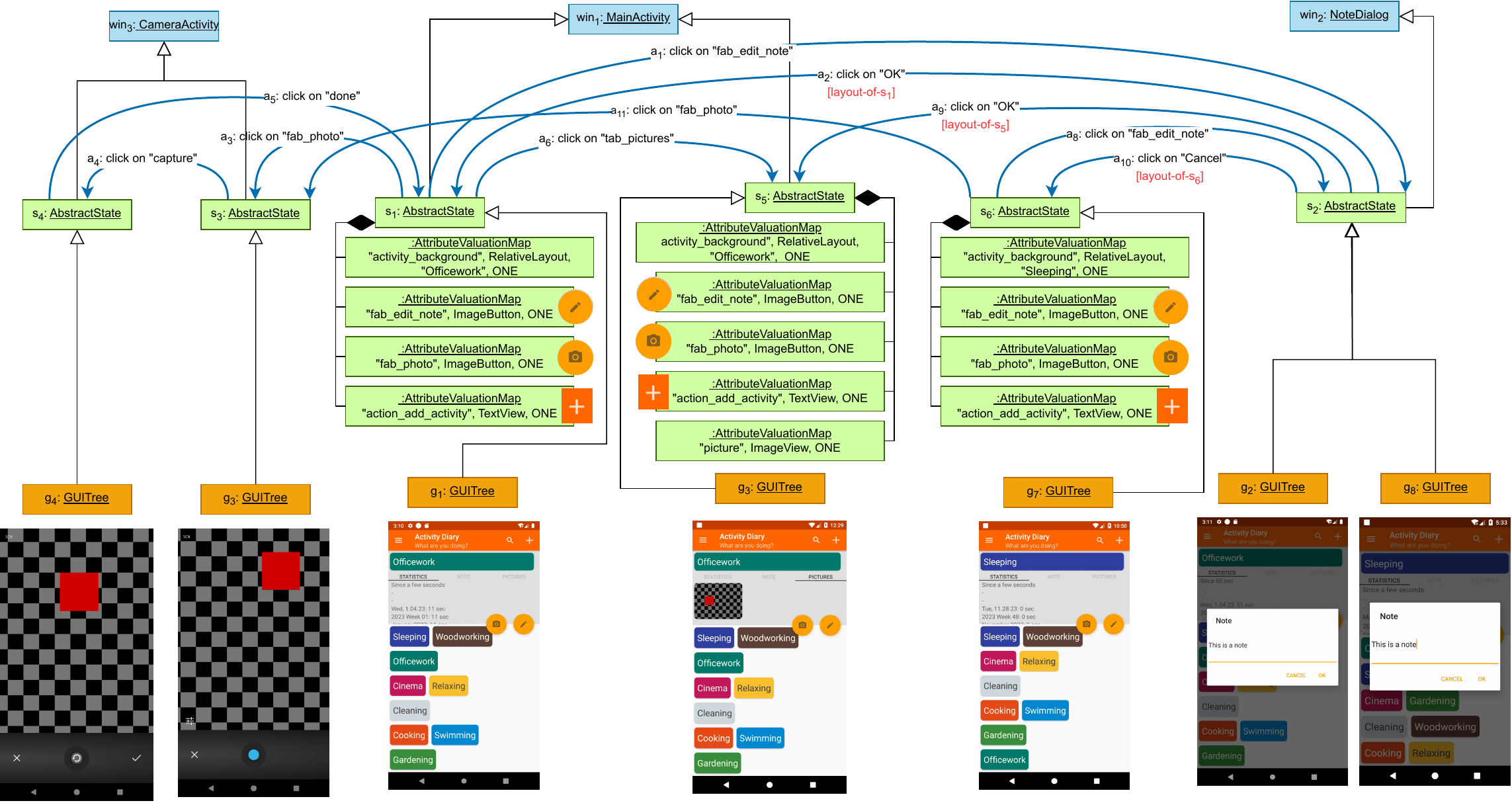}
    
         \begin{footnotesize}
            \textbf{Note:} The dialog window's background is not captured by our \RFs; therefore, the dialogs shown after $g_1$ and $g_3$ belong to the same AbstractState (i.e., $s_2$)
         \end{footnotesize}
         \caption{Closing a dialog brings the App back to the same screen where the dialog was opened.}
         \
         \label{fig:memory-guarded-example1}
\end{figure}

State abstraction may lead to \emph{non-deterministic AbstractTransitions} when the effect of an Action depends on a previously performed Action. \MAJOR{R3.5a}{Recall that \CALM keeps track of the current concrete state of the App, which is modeled as a GUITree in the GSTG (see Figure~\ref{fig:app:model:metamodel:FULL}, Page~\pageref{fig:app:model:metamodel:FULL}). Also, each concrete state is mapped to an AbstractState in the DSTG, and the DSTG is used to determine a sequence of Actions that should trigger the  AbstractTransitions leading to a target AbstractState or Window. 
We observe \emph{non-deterministic AbstractTransitions} when two (or more) AbstractTransitions departing from a same AbstractState are triggered by a same input (e.g., click on the same Button) but lead to two (or more) distinct AbstractStates. 
Although such non-determinism may be due to the App design (e.g., you click on a button and the App always open a different random Window), it is expected for Actions to be reasonably deterministic (e.g., the same Window is opened, although the content may slightly change, as in a news App). Therefore, we assume that non-determinism mainly depends on the state abstraction approach inherited from ATUA. Ideally, we would like to rely on a state abstraction mechanism that minimizes non-determinism to effectively drive testing.} 

\MAJOR{3.5a}{We empirically observed that such non-determinism usually occurs} when AbstractTransitions bring the App into an AbstractState recently visited or into an AbstractState derived from a recently visited one. The latter often consists of an AbstractState having the same layout but a different number of Widgets than the previous AbstractState. In both those scenarios, the previous state may be different at every execution of the AbstractTransition. 
An example based on Activity Diary is shown in  Figure~\ref{fig:memory-guarded-example1}: the action of clicking on the \emph{Close dialog} button in state $s_2$ brings the App to the same screen (and same AbstractState) visualized before opening the dialog (i.e., either $s_1$ or $s_5$).

Note that non-deterministic abstract transitions are more frequent in reused models because \MAJOR{R3.5b}{these models are incrementally built during multiple test sessions (one per App version) and, consequently, they contain several transitions, including non-deterministic ones. Models built from scratch contain no AbstractTransitions when testing starts, CALM incrementally creates AbstractTransitions during testing and, when there are no AbstractTransitions, the test algorithm simply follows WindowTransitions, which do not lead to any expected AbstractState. As a result, the probability of selecting an input sequence that traverses a non-deterministic AbstractTransition is much lower when not reusing models.  Further, non-deterministic AbstractTransitions are more likely to lead to infeasible Action sequences when models are reused because, in this case, testing starts with a populated DSTG although the App was just started. What may happen is that, at the beginning of testing, CALM selects an Action sequence because it is shorter than another one, but the selected sequence is invalid because derived from a non-deterministic transition. ATUA, instead, starting from an App model without a DSTG, is likely to select a longer but feasible sequence of Actions (e.g., because relying the EWTG).} 
For example, in the case of Figure~\ref{fig:memory-guarded-example1}, the App starts in screen $g_7$ of state $s_6$ and \CALM may try to perform the infeasible sequence $\langle a_8,a_9\rangle$ since it is shorter than the feasible sequence $a_{11}, a_4, a_5, a_6$, 
which first adds a photo for the current activity and then opens the "Pictures" tab. 


To handle non-determinism due to AbstractTransitions bringing the App into a recently visited AbstractState or a state derived from a recently visited one, \APPR augments App models with guard conditions specifying if the state reached by the transition is expected to be derived from a previously visited AbstractState. We call such transitions \emph{layout-guarded abstract transitions}. During testing, before adding a state transition to the App model, \CALM verifies 
if the destination state has a layout similar to the layout of any previously visited AbstractState; for that, it processes the GSTG backward till it reaches the initial state or a GUITree with an AbstractState having a layout similar to the destination state. 
To determine if two AbstractStates have a similar layout, \CALM focuses on  the AttributeValuations derived using the reducers belonging to \MAJOR{3.9}{the abstraction function  \RFone,} which does not include text and Widget children; indeed, text and Widget children are likely to vary when AbstractStates are derived from previously visited ones. 
Two AbstractStates have a similar layout if they share 80\% of such AttributeValuations; we propose a threshold of 80\% because it led to the best results (highest coverage of modified methods and instructions belonging to modified methods) in a preliminary experiment conducted with the latest version of all the subject Apps included in our empirical evaluation.


When generating Action sequences, \APPR traverses \emph{layout-guarded} AbstractTransition only if the referenced layout is similar to the layout of an AbstractState  previously visited. 
    In Figure \ref{fig:memory-guarded-example1}, the transition from $s_2$ to $s_5$, triggered by the Action $a_9$, is guarded by $\mathit{layout\_of}(s_5)$. Thanks to such guard, at runtime, \CALM will not suggest the infeasible sequence \MAJOR{R3.10}{$\langle a_8,a_9 \rangle$} when being in $s_6$ . Indeed, the AVMs of $s_5$ and $s_6$ derived with \RFone differ because $s_5$ includes a picture Widget that is not present in $s_6$. 

\subsubsection{Probabilistic Action sequences to handle state explosion} \label{sec:probabilistic-path}

\MAJOR{R3.6a}{In \CALM and \ATUA, 
an AbstractState is modelled as a set of tuples (called AbstractValuationMaps), with each tuple being populated with the values returned by a reducer applied to a Widget visualized on the screen. For that reason, a change in the set of Widgets being visualized (e.g., a button visualized only under certain conditions) affects the set of tuples being generated, and leads to a different state. Consequently, in the presence of several Widgets, that may be hidden or visualized, the number of reachable states may explode.} 

\MAJORBEGIN{}
ATUA and \CALM already include strategies to prevent state explosion due to multiple instances of a same Widget, all fitting a same purpose (e.g., multiple activity buttons in Activity Diary, one for each activity tracked by the App that, when clicked, open the activity details). Indeed, when multiple Widgets lead to a same tuple, \CALM simply indicates that the AbstractValuationMaps has a cardinality above $1$. However, such a solution does not prevent observing different AbstractStates when (A) different sets of Widgets are visualized by a same Window (e.g., a button appearing or disappearing) or when (B) a fine-grained abstraction function (e.g., \RFtwo, which captures the text label of Widgets) is adopted; please note that, in general, deriving different AbstractStates in such cases is correct because (A) the presence of a different set of Widget types indicates that an App is providing a different set of features to the end-user and (B) capturing text labels often prevents non-deterministic transitions.
\MAJOREND{}
%
%
 For example, in the Activity Diary main window (e.g., $g_1$ and $g_3$ in Figure~\ref{fig:memory-guarded-example1}), clicking on the ``Pictures'' tab makes a picture Widget appear and a text Widget disappear, thus separating $g_3$ and $g_7$ into two different AbstractStates, which is desirable because the two states present different testing opportunities (i.e., visualizing text in one case versus a picture in the other).

\begin{figure}[b]
\includegraphics[width=14cm]{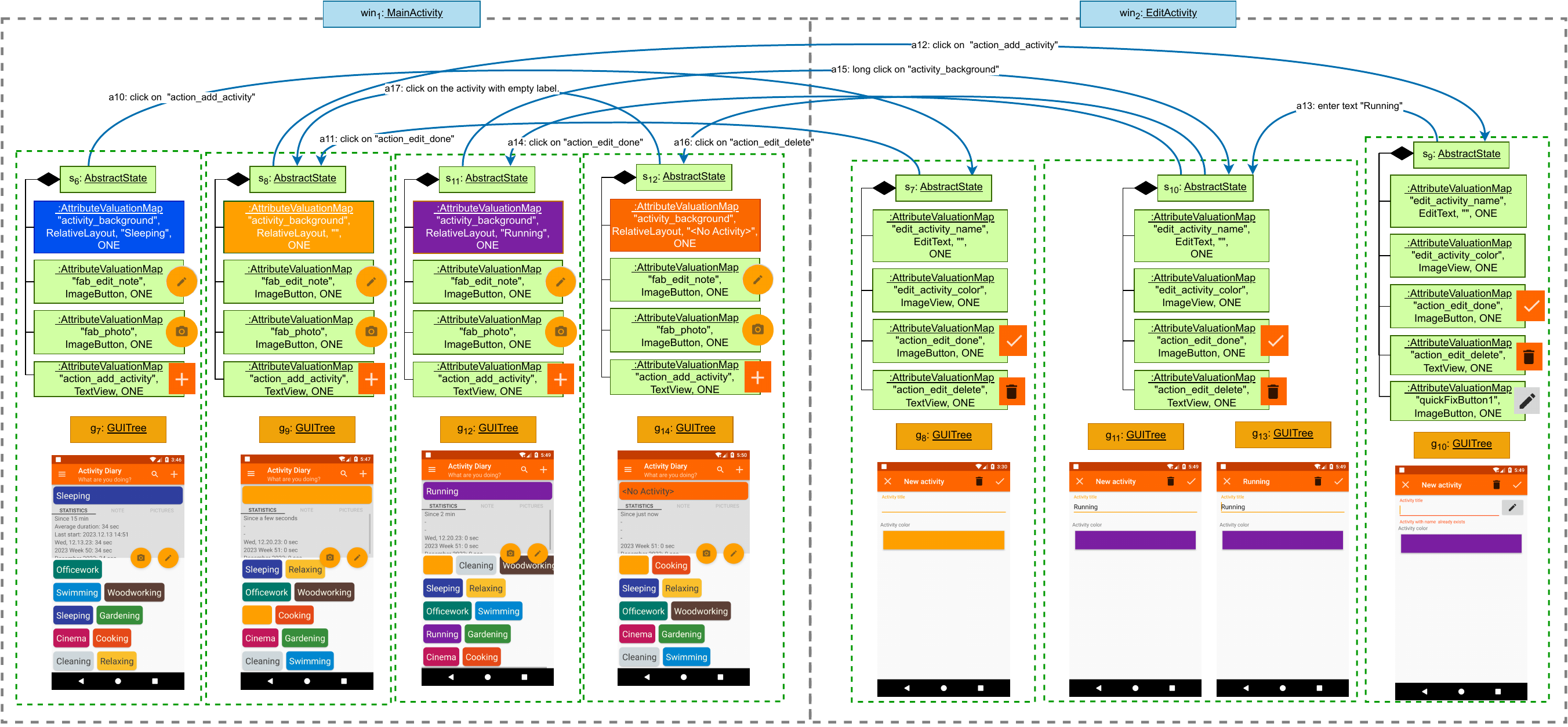}
\caption{DSTG showing the effect of state explosion.}
\label{fig:state:explosion:example}
\end{figure}

 \MAJOR{R3.6b}{An example of how the ATUA/CALM abstraction function may lead to state explosion is shown in Figure~\ref{fig:state:explosion:example}, where \RFtwo, which takes into account the text of a widget to distinguish abstract states, creates different AbstractStates for each App screen with a different subset of activities (e.g., ``Sleeping'' and ``Woodworking'' versus ``Sleeping'', ``Woodworking'', and ``Officework'' ); indeed, the GUITrees $g7$, $g9$, $g12$, and $g14$ differ only with respect to the text being displayed in the top part of the App screen but lead to four different AbstractStates (i.e., $s_6$, $s_8$, $s_{11}$, and $s_{12}$). 
 With \RFtwo, if the App is in the GUITree $g_7$ (AbstractState $s_6$, shown in Figure~\ref{fig:memory-guarded-example1}), ATUA and CALM derive the DSTG in Figure~\ref{fig:state:explosion:example} by executing the following actions:
 (1) adding a new activity (i.e., clicking on the ``+'' button, $a_{10}$), which brings the App to the ``Edit Activity Window'' in $g_8$ and AbstractState $s_7$, (2) validating the activity without giving it a name (i.e., clicking on the ``V'' button, $a_{11}$, which brings the App to $g_9$ and AbstractState $s_8$), (3) adding another new activity ($a_{12}$), which leads to a new AbstractState ($s_9$) where the button to fix the activity name (i.e., ``quickFixButton1'') appears, (4) entering some text in the activity name (i.e., ``Running'', $a_{13}$), which leads to $s_{10}$ because the text area is not empty anymore, (5) validating the activity ($a_{14}$), which leads to $g_{12}$ and AbstractState $s_{11}$, (6) long clicking on the ``activity\_background'' labeled ``Running'' ($a_{15}$), which leads to $g_{12}$ (state $s_{10}$), (7) deleting the activity ($a_{16}$), which leads to a new AbstractState ($s_{12}$) where we observe the same labels as state $s_{8}$ but there is no activity selected, and finally (8) clicking on the activity with the empty label ($a_{17}$), which leads to $s_8$.} 


 \MAJOR{R3.6b}{Such state explosion has a detrimental effect on test effectiveness because, although all the AbstractStates of the \emph{Main Activity} Window may bring the App into a target AbstractState with a same number of actions, ATUA and \CALM can determine that an AbstractState $s_t$ reaches a target AbstractState $s_z$ only if there is an AbstractTransition connecting $s_t$ and $s_z$, which happens only if CALM has already exercised the Input bringing $s_t$ into $s_z$ in the past, an unlikely condition with state explosion. 
 For example, in Figure~\ref{fig:state:explosion:example}, although the AbstractState $s_9$ can be reached by clicking the ``plus'' button in the AbstractStates $s_8$, $s_{11}$, and $s_{12}$, 
 ATUA cannot know that state $s_{12}$ can reach state $s_9$, and  thus, if the current state is $s_{12}$, ATUA, instead of clicking on the ``plus'' button, will first try to reach an AbstractState with an AbstractTransition going into $s_9$ (e.g., $s_8$).}

 \MAJOR{R3.6c}{The consequences of state explosion are exacerbated in the case of model reuse. Specifically, to reach target AbstractStates  that in previous testing sessions have been observed at a late stage (e.g., after testing other features), ATUA would suggest an Action sequence that traverses all the AbstractStates observed,  even if a shorter Action sequence would be feasible. 
 For example, let us assume that, when testing previous versions of Activity Diary, ATUA has derived the DSTG in Figure~\ref{fig:state:explosion:example}. Also, assume that the current state is $s_{11}$ and ATUA needs to test the method that suggests a new name when the provided name for a new activity already exists, which means that ATUA should reach the AbstractState $s_9$ (Figure~\ref{fig:state:explosion:example}) and then click on ``quickFixButton1''.
 Since there is no transition from $s_{11}$ to $s_9$, ATUA would 
   suggest to perform the following four-action sequence: long click on ``activity\_background'' ($a_{15})$, click on ``action\_edit\_delete'' ($a_{16}$), click on the activity with empty label ($a_{17}$),  click on ``action\_add\_activity'' ($a_{12}$). However, such suggestion is sub-optimal because clicking on ``action\_add\_activity'' ($a_{12}$) brings the App into $s_9$ directly from $s_{11}$.} 
   
 


\begin{figure}[b]
\includegraphics[width=12cm]{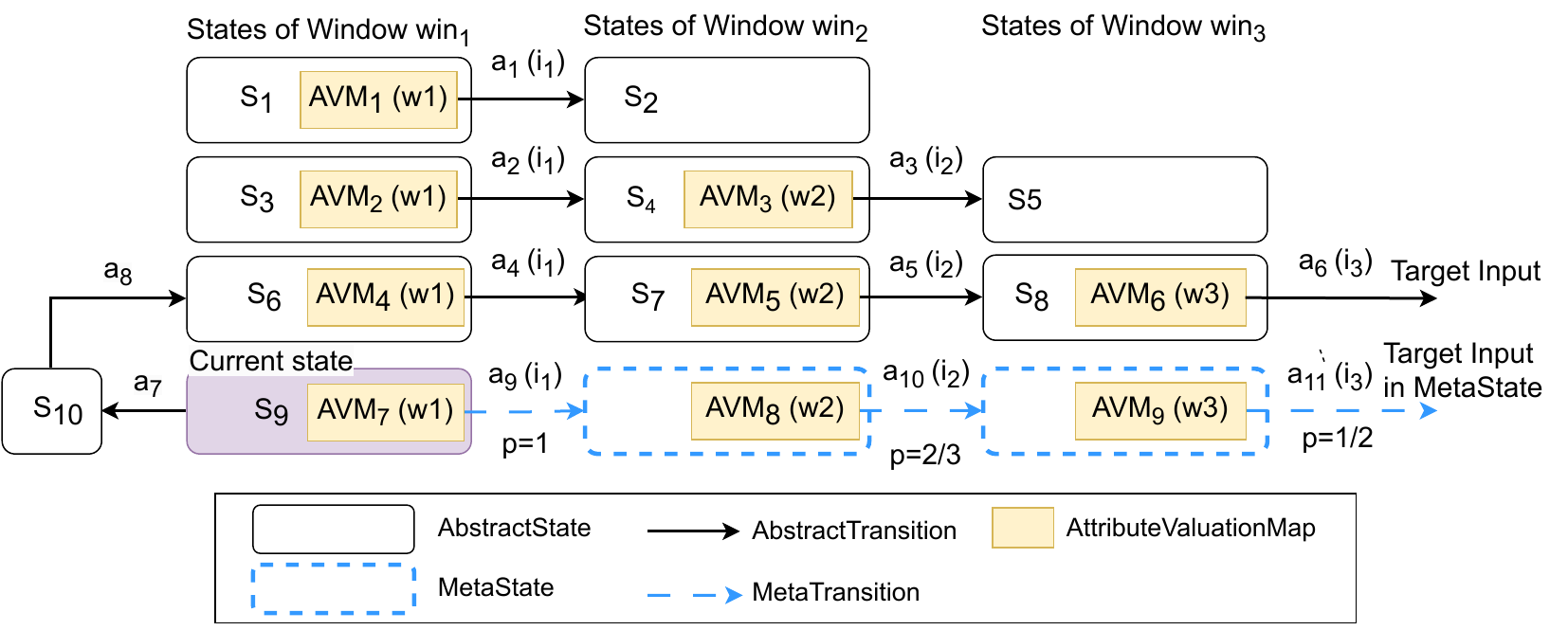}
\caption{Part of a DSTG with a Probabilistic Action sequence; MetaStates and MetaTransitions are dashed.}
\label{fig:PAsequence}
\end{figure}

In practice, in the presence of \emph{state explosion due to dynamically appearing widgets}, 
a sufficient condition to execute a whole Action sequence 
is often the presence of all the Widgets targeted by the Actions in the sequence. Therefore, an Action sequence should be selected by identifying the Actions leading to App screens that likely contain the Widgets targeted by the next Action in the sequence, 
till a Window with the targeted Input is reached. 
In Figure~\ref{fig:PAsequence}, we provide an abstract example resembling what reported in Figure~\ref{fig:state:explosion:example}, with a current state ($s_9$) where the shortest Action sequence reaching the target input (i.e., the sequence $\langle$\emph{click on $w1$, click on $w2$, click on $w3$}$\rangle$ derived from the AbstractState $S_6$) cannot be derived from the DSTG although it might be feasible even if $S_9$ is observed.
\MAJOREND{}

Because of the observation above, \CALM derives \emph{probabilistic Action sequences} in addition to \emph{deterministic Action sequences} (i.e., what is derived by \ATUA). 
\MAJOR{3.13}{For every Action, they capture the
probability that the App visualizes the Widget required to perform the action; in other words, they capture the probability for each action to be performed.
Thanks to such probabilistic approach, they enable constructing Action sequences with Actions that exercise AbstractTransitions that have not been exercised in the past, but are likely feasible (i.e., \CALM never exercised the input triggering a specific AbstractTransition, but it can estimate what Widgets will be visualized next, based on previous executions). 
A DSTG like the one in Figure~\ref{fig:PAsequence} enables the derivation of the sequence $\langle$\emph{$a_9$, $a_{10}$, $a_{11}$}$\rangle$ although there is no AbstractTransition departing\footnote{Please recall that AbstractTransitions depart from the AVM of a specific Widget in a given AbstractState.} from the $\text{AVM}_7$ of widget $w_1$ in $S_9$.}

\MAJOR{R3.11}{Since \CALM traverses the DSTG to derive Action sequences, in order to obtain such probabilistic Action sequences, it is necessary to generate a DSTG that includes also AbstractTransitions that have never been triggered previously, but are likely to be feasible. We call such additional transitions \emph{MetaTransitions}. 
\CALM incrementally adds MetaTransitions to the App model while traversing the DSTG to derive an Action sequence. Specifically, \APPR adds a MetaTransitions for every Input $i$ that has never been exercised in the current AbstractState. After selecting an Action sequence, to avoid looking for outdated MetaTransitions and updating them, \CALM discards all the MetaTransitions added to the App model.}

The destination of a MetaTransition is a \emph{MetaState}; a MetaState tracks all the Widgets ever visualized in any App screen after exercising the Input $i$. To derive MetaStates,  \APPR considers the AbstractStates reached after exercising $i$ and identifies all the Widgets belonging to them.
\MAJOR{3.13}{If an Input $i$ may lead to distinct Windows (i.e., it is exercised by non-deterministic AbstractTransitions leading to different Windows), \CALM will generate one MetaState for each Window reached through Input $i$ and, therefore, will add multiple MetaTransitions departing from the same AVM (one for each MetaState to be reached).
Each MetaTransition leaving a MetaState is associated with a probability (\emph{MetaTransition probability}) of being available; the MetaTransition probability captures the likelihood that the Widget targeted by the MetaTransition is interactable in any of the observed AbstractStates and is computed as follows:
$$p_i=\frac{\text{number of AbstractStates of } \mathit{win}_{i} \text{ including } w_{i}}{\text{number of AbstractStates of } \mathit{win}_{i}}$$
with $\mathit{win}_{i}$ being the Window to which Input $i$ belongs, and $w_{i}$ being the Widget that is exercised by Input $i$.}

In Figure~\ref{fig:PAsequence},
Input $i_{1}$, which targets Widget $w_1$ in Window $\mathit{win}_1$, may bring the App into three distinct AbstractStates, two of which including widget $w_2$; the probability of exercising $w_1$ is 1.0 ($w_1$ is present in the current state $S_9$), the probability of reaching $w_{2}$ after exercising $i_{1}$ is $0.67$ (i.e., $2/3$) and therefore the MetaTransition probability for $a_{10}$ is $0.67$. Similarly, the probability of exercising the target input on $w_3$ is $0.5$.

During testing, \APPR should select the Action sequence that brings the App into the target Window/AbstractState with a minimal cost.
However,
MetaTransitions may not bring the App into a desired MetaState, and thus it may not be feasible to fully exercise a probabilistic Action sequence. To maximize efficiency, we should estimate an Action sequence cost by accounting for the risk of not reaching a desired MetaState.


We assume that \emph{deterministic Action sequences}
are likely to be fully executed because they do not include MetaTransitions; therefore, their cost depends on the time required to execute all their Actions.
Given a \emph{probabilistic Action sequence}  and a  \emph{deterministic Action sequence} with the same length, \CALM should exercise the deterministic one because it does not present feasibility risks\MAJOR{R3.14}{; consequently, the cost computed for the probabilistic action sequence should be higher. To satisfy such property, we define a cost function that sums the cost of executing the whole sequence $\phi$ with the cost of executing a sequence $\phi$ that is infeasible:} 
\MAJORBEGIN{}
$${ cost(\phi)= cost\_full(\phi) + infeasibility\_cost(\phi) }$$

 The cost of executing a whole Action sequence $\phi$ (either partial or deterministic) is the sum of the cost of executing all its $n$ Actions:

$${cost\_full(\phi) =
\sum_{j=1}^{n} cost(\mathit{action}_{j})}$$

Since we empirically observed that all the Actions, except App reset, take almost the same time to execute, and App reset takes around ten times the other Actions, we assign to reset Actions a cost of $10$, and $1$ to all other Actions. 

The infeasibility cost, instead, depends on the likelihood of executing a sequence only partially (given two \emph{probabilistic Action sequences} with the same length, \CALM should exercise the one that is more likely to be feasible) and on the cost of executing only a subset of Actions. The latter is due to \CALM having to recompute, when an Action sequence is partially executed, another Action sequence to reach the target AbstractState/Window; in practice, partially executed Action sequences lead to a waste of time 
(e.g., \CALM may need to go back to the previous AbstractState).

Because of the above, we heuristically compute such infeasibility cost as the cost of 
executing a partial subsequence of $\phi$ without getting closer to the target AbstractState/Window, multiplied by the likelihood of not executing the whole sequence $\phi$:\\


$${ infeasibility\_cost(\phi)= 
cost\_partial(\phi) 
*
likelihood\_partial(\phi)}$$

\MAJOREND{}

\MAJOR{R3.15}{The cost of executing only part of an Action sequence depends on the number of Actions being triggered and on the likelihood that such Actions bring the App into an AbstractState that is not closer to the target AbstractState/Window (otherwise the cost is mitigated because the next Action sequence will be shorter). Since we cannot predict how many executed Actions help reach the target AbstractState/Window, we conservatively assume that, on average, half of the Actions belonging to a \emph{probabilistic Action sequence} lead to the desired state.} 

\[ 
{
cost\_partial(\phi) =
\frac{\sum_{j=1}^{n} cost(action_{j})}{2}
}\]

Finally, the likelihood of executing an input sequence depends on the likelihood of observing all the Widgets targeted by the Actions in the sequence. Therefore, 
%
the probability of partially executing an input sequence $\phi$ is:

$$
{likelihood\_partial(\phi) = 1 - \prod_{j=1}^{n} p(\mathit{action}_{j}|s_{j-1})}$$

with $p(\mathit{action}_{j}|s_{j-1})$ being the probability of observing the Widget required to trigger the Action $\mathit{action}_j$ in the state reached by $\mathit{action}_{j-1}$.
 If $s_{j-1}$ is not a MetaState, we expect the widget to be available; therefore,  $p(\mathit{action}_{j}|s_{j-1}) = 1.0$. 
 If $s$ is a MetaState, $p(\mathit{action}_{j}|s_{j-1})$ matches the \emph{MetaTransition probability} described above. 
For a deterministic input sequence $\phi_d$, since input widgets are available in the reached AbstractStates (i.e., $p(i_{j}|s_{j-1}) = 1$), we have:

$${likelihood\_partial(\phi_d) = 0}$$ 

 During testing, when executing a partial input sequence $\phi$, for each  MetaState $s_e$ expected after an Action $i_j$, \APPR verifies that the Widget to be triggered next is available; otherwise a new input sequence needs to be selected. 

Based on the above, in the example in Figure~\ref{fig:PAsequence}, the probabilistic Action sequence
$\langle a_9, a_{10}, a_{11} \rangle$ will be selected instead of the deterministic Action sequence $\langle a_7,a_{8},a_4, a_5, a_6 \rangle$,  thus reducing the number of Actions required to reach the test target.
Indeed, since the current state is $S_9$ and the objective is to reach $w3$ and trigger $i_3$, the cost of the probabilistic action sequence $\langle a_9, a_{10}, a_{11} \rangle$ would be computed according to the following equations:

\begin{equation}
\left\{\begin{aligned}
cost(\langle a_9, a_{10}, a_{11} \rangle)&= cost\_full(\langle a_9, a_{10}, a_{11} \rangle) \\
&+
cost\_partial(\langle a_9, a_{10}, a_{11} \rangle)*likelihood\_partial(\langle a_9, a_{10}, a_{11} \rangle)\\
cost\_full(\langle a_9, a_{10}, a_{11} \rangle)&=cost(a_9)+cost(a_{10})+cost(a_{11})=1+1+1=3\\
cost\_partial(\langle a_9, a_{10}, a_{11} \rangle)&=\frac{cost(a_9)+cost(a_{10})+cost(a_{11})}{2}=1.5\\
likelihood\_partial(\langle a_9, a_{10}, a_{11} \rangle)
&=1-\big(p(a_9)*p(a_{10})*p(a_{11})\big)\\
&=1-(1*2/3*1/2)=0.66
\end{aligned}\right.
\end{equation}

They lead to: 

\begin{equation}
\begin{aligned}
cost(\langle a_9, a_{10}, a_{11} \rangle)&=3+(1.5*0.66)=3.99
\end{aligned}
\end{equation}

The Action sequence $\langle a_7, a_8, a_4, a_5, a_6 \rangle$, instead, leads to:

\begin{equation}
\begin{aligned}
cost(\langle a_7, a_8, a_4, a_5, a_6 \rangle)&=
cost\_full(\langle a_7, a_8, a_4, a_5, a_6 \rangle)\\
&+ 
cost\_partial(\langle a_7, a_8, a_4, a_5, a_6 \rangle) * likelyhood\_partial(\langle a_7, a_8, a_4, a_5, a_6 \rangle)\\
&=\big((cost(a_7)+cost(a_{8})+cost(a_4)+cost(a_5)+cost(a_6)\big)\\
&+ 
cost\_partial(\langle a_7, a_8, a_4, a_5, a_6 \rangle) * likelyhood\_partial(\langle a_7, a_8, a_4, a_5, a_6 \rangle)\\
&=5+cost\_partial(\langle a_7, a_8, a_4, a_5, a_6 \rangle) * 0\\
&=5
\end{aligned}
\end{equation}

\subsubsection{Backward-equivalent abstract states detection}
\label{sec:runtime:adapt}

\MAJOR{2.6}{In the presence of modified Windows, AbstractStates may not match across versions although they are the source of AbstractTransitions that behave the same across versions. 
We call such AbstractStates \emph{backward-equivalent AbstractStates}; they are observed in the presence of minimal changes in App Windows (e.g., few Widgets being added or removed). 
For example, if the updated App introduces a reset button for a Window with a form, an Action sequence exercising the submit button, which has not been modified, should remain executable in the updated version; however, the AbstractState of the two Window versions does not match because of their different number of Widgets.}

A \emph{backward-equivalent AbstractState} differs from an AbstractState expected by the input sequence and inherited from the base App's DSTG, but enables performing the same actions.
Precisely, an AbstractState $s_o$ observed in the updated App is backward-equivalent to a state $s_e$ derived from the base DSTG when:

\begin{itemize}[leftmargin=*]
    \item $s_e$ and $s_o$ are associated to the same Window; otherwise, they cannot be equivalent because different Windows implement different features.
    \item Every AVM in $s_e$  
    matches an AVM in $s_o$. Otherwise, it would not be possible to trigger, in $s_o$, the same Actions triggerable in $s_e$.
    \item Every AVM in $s_o$, 
    except the ones for the EWTG Widgets added or replaced in the updated App,
    matches an AVM in $s_e$. 
    If this condition does not hold, a Widget may have different AVMs in the two App versions (e.g., a checkbox is no longer checked). In such case the updated App changed its behaviour and, consequently, a same Action may not exercise the same methods in the base and updated version. 
\end{itemize}

For example, taking the AbstractState $s_2$ in Figure~\ref{fig:step2-example}, which is inherited from the base App Mode, as an expected AbstractState, CALM observes a new AbstractState $s_3$ similar to $s_2$ but including additionally an Attribute ValuationMap representing the added EWTG Widget $w_9$ in the "EditActivity" Window. Since there is only a mismatch between $s_3$ and $s_2$, which is related to the added EWTG Widget, $s_3$ is backward-equivalent to $s_2$. This implies that any required action that needs to be performed on $s_2$ could be done on $s_3$ too.

\MAJOR{R3.16}{During testing, when \CALM encounters an AbstractState that does not match the expected one, it verifies if it is backward equivalent. If such state is backward equivalent, \CALM proceeds with executing the rest of the current Action sequence and otherwise discards the current Action sequence and generates a new one.}


\subsubsection{Online App model refinement to deal with obsolete abstract states}
\label{sec:online:refinement}

\MAJOR{2.6}{We may observe \emph{obsolete AbstractStates}. 
AbstractStates often depend on a remote component (e.g., a news server) that provides data that change over time. Consequently, such AbstractStates become obsolete at run-time (e.g., after a few minutes) or in the time span between two App versions. Other AbstractStates become obsolete because the behaviour of the updated App changed (i.e., it is not possible to reach a certain AbstractState with the same input sequence observed in a previous App version).}

Online model refinement aims at determining if expected states that are not observed when exercising an Action sequence are obsolete. 
It is necessary because, otherwise, \CALM may keep selecting Action sequences that include an unreachable state, which leads to a waste of resources.

When exercising an Action sequence, if the state expected after the $i^{th}$ action ($s_{e_i}$) does not match (or is backward-equivalent to) the observed state ($s_{o_i}$), \CALM determines if  $s_{e_i}$ is obsolete. 
If $s_{e_i}$ has already been observed when testing the updated App, then it is not obsolete; therefore, $s_{o_i}$ is the result of nondeterminism and \APPR applies ATUA's procedure to minimize nondeterminism (i.e., it refines $s_{e_{i-1}}$ using \RF).
Otherwise (i.e., if $s_{e_i}$ had not been observed with the updated App), \APPR  removes from the DSTG the AbstractTransition connecting $s_{e_{i-1}}$  with $s_{e_i}$.

 \MAJOR{3.7}{Note that, at a high level, all the limitations above  depend on the specific state abstraction strategy adopted in CALM. If CALM relied on a coarse-grained state abstraction strategy not taking into consideration the Widgets in a Window to derive an AbstractState (e.g., each Window can have only one AbstractState), we would not have observed any of those limitations, except for non-deterministic AbstractTransitions. Indeed, such coarse-grained strategy would not capture any difference in the Widgets present in the Windows visualized by two App versions, thus not leading to obsolete states or states not matching but rather backward-equivalent states; similarly, it would not lead to state explosion. However, such state abstraction strategy would likely lead to several non-deterministic transitions (e.g., not distinguishing between submitting an empty and a filled form) thus leading to  ineffective testing. A finer-grained strategy (e.g., relying on the GSTG), instead, would prevent non-deterministic transitions, but would favor state explosion.}

\subsection{Step 4: Refine the App model offline.}


After testing an App version $V_x$, \emph{to further clean up the App model from 
unreachable AbstractStates},  \APPR
removes from the App model all those AbstractStates that were not visualized although belonging to exercised Windows. 

Further, \emph{to remove AbstractStates that become quickly obsolete}, 
after testing an App version $V_x$, \APPR re-executes, offline, the sequences of test inputs captured by the GSTG. 
During such re-execution, if a test input does not bring the App into the expected AbstractState $s_e$, then \APPR annotates $s_e$ as obsolete. When testing  version $V_{x+1}$, to avoid wasting the test budget, \APPR does not generate input sequences that traverse obsolete states. 

Finally, since we empirically observed that Windows with obsolete states often present newer states that quickly become obsolete (e.g., Apps that display updated news every few minutes),
\APPR considers obsolete any AbstractState identified when testing $V_{x+1}$ but not reachable with Action sequences traversing their incoming transitions, while still testing $V_{x+1}$.
Our approach enables \APPR to rely on obsolescent AbstractStates till they become obsolete.





\section{Empirical Evaluation}
\label{sec:empirical} 

We performed an empirical evaluation that aims to address the following research questions (RQs):

\begin{itemize}
\MAJORBEGIN{}
 \item \MAJOR{3.17}{\emph{RQ1.} \emph{Does \CALM preserve the capability of ATUA to reduce test oracle costs?} 
Since in App testing, each input leads to one App screen to be visually inspected, minimizing the number of inputs reduces testing cost, as discussed in previous work~\cite{ATUA}. We aim to study if model reuse preserves the main advantage of ATUA, which was demonstrated to be the best approach to minimize the number of exercised inputs~\cite{ATUA}.}
\MAJOREND{}
\item \emph{RQ2.} \emph{Is \APPR more effective than competing approaches in testing App updates, for a same test budget?} We aim to determine if \APPR performs significantly better (code coverage) than ATUA and SOTA approaches that complement ATUA~\cite{ATUA}. 
\item \emph{RQ3.}  \emph{How do \APPR and competing approaches fare, for different testing time budgets, with updates of different magnitude?} 
    Updated Apps may need to be tested quickly (e.g., after each code commit, with a limited test budget).  
    However, both the magnitude of the update (e.g., number of updated methods) and the testing time budget may affect the performance of \CALM. 
    Therefore, we study how the effectiveness of \CALM compares with competing approaches over time and for updates of different magnitude.
   \MAJORBEGIN{}
   \item \emph{RQ4.} \emph{To what extent does \CALM enable engineers to detect functional faults?}
    We aim to evaluate the effectiveness of \CALM in generating input sequences that exercise faults and report output screens showing failures. 
   \MAJOREND{}
\end{itemize}

Our replication package is available online~\cite{replicability}.

\subsection{Subjects of the study}

Since \APPR extends ATUA, 
we reuse all the subjects used for evaluating ATUA, except those that cannot be tested anymore because relying on dismissed server-side APIs.   

Table~\ref{table:caseStudies} shows the selected versions (52 subjects, in total);
Table~\ref{table:updated} provides the number of updated methods for each version; they
range from one (version V7 of Activity Diary) to 603 (BBC's V5),
thus being representative of diverse release scenarios (i.e., from bug fixes to
major releases). The number of bytecode instructions shows that our subjects vary in complexity (from 3667 to 163303). 



\begin{table}[tb]
\caption{Selected subject systems.}
\label{table:caseStudies}
\scriptsize
\begin{tabular}{|p{3mm}|
>{\raggedleft\arraybackslash}p{6mm}@{\hspace{1pt}}|
@{\hspace{0pt}}>{\raggedleft\arraybackslash}p{6mm}@{\hspace{1pt}}|
@{\hspace{0pt}}>{\raggedleft\arraybackslash}p{6mm}@{\hspace{1pt}}|
@{\hspace{0pt}}>{\raggedleft\arraybackslash}p{6mm}@{\hspace{1pt}}|
@{\hspace{0pt}}>{\raggedleft\arraybackslash}p{6mm}@{\hspace{1pt}}|
@{\hspace{0pt}}>{\raggedleft\arraybackslash}p{6mm}@{\hspace{1pt}}|
@{\hspace{0pt}}>{\raggedleft\arraybackslash}p{6mm}@{\hspace{1pt}}|
@{\hspace{0pt}}>{\raggedleft\arraybackslash}p{6mm}@{\hspace{1pt}}|
@{\hspace{0pt}}>{\raggedleft\arraybackslash}p{6mm}@{\hspace{1pt}}|
@{\hspace{0pt}}>{\raggedleft\arraybackslash}p{6mm}@{\hspace{1pt}}|
}
\hline
\multicolumn{1}{|c|}{\textbf{App}}
&\multicolumn{1}{@{\hspace{1pt}}
>{\centering\arraybackslash}p{6mm}@{\hspace{1pt}}|}{\textbf{V0}}
&\multicolumn{1}{@{\hspace{1pt}}
>{\centering\arraybackslash}p{6mm}@{\hspace{1pt}}|}{\textbf{V1}}
&\multicolumn{1}{@{\hspace{1pt}}
>{\centering\arraybackslash}p{6mm}@{\hspace{1pt}}|}{\textbf{V2}}
&\multicolumn{1}{@{\hspace{1pt}}
>{\centering\arraybackslash}p{6mm}@{\hspace{1pt}}|}{\textbf{V3}}
&\multicolumn{1}{@{\hspace{1pt}}
>{\centering\arraybackslash}p{6mm}@{\hspace{1pt}}|}{\textbf{V4}}
&\multicolumn{1}{@{\hspace{1pt}}
>{\centering\arraybackslash}p{6mm}@{\hspace{1pt}}|}{\textbf{V5}}
&\multicolumn{1}{@{\hspace{1pt}}
>{\centering\arraybackslash}p{6mm}@{\hspace{1pt}}|}{\textbf{V6}}
&\multicolumn{1}{@{\hspace{1pt}}
>{\centering\arraybackslash}p{6mm}@{\hspace{1pt}}|}{\textbf{V7}}
&\multicolumn{1}{@{\hspace{1pt}}
>{\centering\arraybackslash}p{6mm}@{\hspace{1pt}}|}{\textbf{V8}}
&\multicolumn{1}{@{\hspace{1pt}}
>{\centering\arraybackslash}p{6mm}@{\hspace{1pt}}|}{\textbf{V9}}
\\

\hline
AD
&105& 	111& 	115& 	118& 	122& 	125& 	130& 	131& 	134 &\\
\hline
BM
&5.1.0& 		5.10.0& 	5.11.0& 	5.12.0& 	5.13.0& 	5.4.0& 	5.5.0& 	5.6.0& 	5.8.1& 	5.9.0\\
\hline
CM
&9.1& 		9.2& 	9.3& 	9.4& 	9.5& 	9.6& 	9.7& 	9.8& 	9.9& 	10.0 \\
\hline
FM
&44& 	53& 	77& 	79& 	82&\multicolumn{5}{c|}{}\\
\hline
WI								
&198& 10239& 10263& 10264& 10269 &\multicolumn{5}{c|}{}\\
\hline
VP
&3.1.4& 		3.1.5& 	3.1.7& 	3.2.12& 	3.2.2& 	3.2.3& 	3.2.6& 	3.2.7& 	3.2.9&\multicolumn{1}{c|}{}\\
\hline
YM
&1.16.0& 	1.16.1& 	1.16.2& 	1.17.3& 	1.18.1& 	1.19.1& 	1.20.1& 	1.20.3& 	1.20.5& 	1.20.7 \\
\hline

\end{tabular}

\footnotesize{
\textbf{Apps:} 
AD: Activity Diary,
BBC: BBC Mobile,
CM: Citymapper,
FM: Amaze File Manager,
WI: Wikipedia,
VP: VLC Player,
YM: Yahooweather Mobile}

\end{table}%

\begin{table}[tb]
\caption{Number of updated methods for each App version. For V0 (assumed as the initial version), we report all the methods of the App.}
\label{table:updated}
\scriptsize
\begin{tabular}{|p{3mm}|
>{\raggedleft\arraybackslash}p{6mm}@{\hspace{1pt}}|
@{\hspace{0pt}}>{\raggedleft\arraybackslash}p{6mm}@{\hspace{1pt}}|
@{\hspace{0pt}}>{\raggedleft\arraybackslash}p{6mm}@{\hspace{1pt}}|
@{\hspace{0pt}}>{\raggedleft\arraybackslash}p{6mm}@{\hspace{1pt}}|
@{\hspace{0pt}}>{\raggedleft\arraybackslash}p{6mm}@{\hspace{1pt}}|
@{\hspace{0pt}}>{\raggedleft\arraybackslash}p{6mm}@{\hspace{1pt}}|
@{\hspace{0pt}}>{\raggedleft\arraybackslash}p{6mm}@{\hspace{1pt}}|
@{\hspace{0pt}}>{\raggedleft\arraybackslash}p{6mm}@{\hspace{1pt}}|
@{\hspace{0pt}}>{\raggedleft\arraybackslash}p{6mm}@{\hspace{1pt}}|
@{\hspace{0pt}}>{\raggedleft\arraybackslash}p{6mm}@{\hspace{1pt}}|
}
\hline
\multicolumn{1}{|c|}{\textbf{App}}
&\multicolumn{1}{@{\hspace{1pt}}
>{\centering\arraybackslash}p{6mm}@{\hspace{1pt}}|}{\textbf{V0}}
&\multicolumn{1}{@{\hspace{1pt}}
>{\centering\arraybackslash}p{6mm}@{\hspace{1pt}}|}{\textbf{V1}}
&\multicolumn{1}{@{\hspace{1pt}}
>{\centering\arraybackslash}p{6mm}@{\hspace{1pt}}|}{\textbf{V2}}
&\multicolumn{1}{@{\hspace{1pt}}
>{\centering\arraybackslash}p{6mm}@{\hspace{1pt}}|}{\textbf{V3}}
&\multicolumn{1}{@{\hspace{1pt}}
>{\centering\arraybackslash}p{6mm}@{\hspace{1pt}}|}{\textbf{V4}}
&\multicolumn{1}{@{\hspace{1pt}}
>{\centering\arraybackslash}p{6mm}@{\hspace{1pt}}|}{\textbf{V5}}
&\multicolumn{1}{@{\hspace{1pt}}
>{\centering\arraybackslash}p{6mm}@{\hspace{1pt}}|}{\textbf{V6}}
&\multicolumn{1}{@{\hspace{1pt}}
>{\centering\arraybackslash}p{6mm}@{\hspace{1pt}}|}{\textbf{V7}}
&\multicolumn{1}{@{\hspace{1pt}}
>{\centering\arraybackslash}p{6mm}@{\hspace{1pt}}|}{\textbf{V8}}
&\multicolumn{1}{@{\hspace{1pt}}
>{\centering\arraybackslash}p{6mm}@{\hspace{1pt}}|}{\textbf{V9}}
\\

\hline
AD
&260& 	18& 	3& 	12& 	117& 	39& 	28& 	1& 	49&\\
\hline
BM
&10706& 	649& 	27& 	44& 	25& 	603& 	242& 	553& 	77& 	95\\
\hline
CM
&9629& 	51& 	37& 	55& 	73& 	119& 	76& 	73& 	12& 	69\\
\hline
FM
&2042& 	306& 	415& 	11& 	644&\multicolumn{5}{c|}{}\\
\hline
WI								
&7477& 	1430& 	535& 	13& 	94&\multicolumn{5}{c|}{}\\
\hline
VP
&6796& 	672& 	26& 	3& 	149& 	13& 	51& 	33& 	42&\multicolumn{1}{c|}{}\\
\hline
YM
&2932&  5&   4& 	243& 	10& 	16& 	118& 	101& 	12& 	9 \\
\hline

\end{tabular}

\footnotesize{
\textbf{Apps:} 
AD: Activity Diary,
BBC: BBC Mobile,
CM: Citymapper,
FM: Amaze File Manager,
WI: Wikipedia,
VP: VLC Player,
YM: Yahooweather Mobile}
\end{table}%

\subsection{Experiment setup}

We compare \APPR with ATUA and five SOTA tools:  APE\cite{Gu:APE:ICSE:2019}, TimeMachine~\cite{TimeMachine}, Monkey\cite{monkey}, Fastbot2~\cite{Fastbot2}, and Humanoid~\cite{Humanoid}. APE is the SOTA tool that is more likely to achieve the highest coverage for a one-hour test-budget~\cite{combodroid}. Monkey, which employs a pure random testing strategy, is the de-facto standard baseline used in the literature~\cite{Wang:EmpStudy:2018, combodroid}. TimeMachine improves over Monkey by leveraging emulators' snapshots to keep a pool of interesting App states (i.e., reached after improving coverage) to resume testing from, when coverage improvement gets stuck.
Fastbot2 is a recent approach reusing App models across versions. Humanoid relies on deep learning to effectively exercise Apps like humans~\cite{Humanoid}. \MAJOR{R2.2}{In the case of Fastbot2, we configured it to reuse models built when testing previous App versions.}
Further, to perform an ablation study, we implemented a version of \ATUA (ATUA-R) that reuses models across versions (i.e., implements \APPR's Steps 1 and 2 but not the heuristics of Step 3); also, we implemented a version of TimeMachine (i.e., TimeMachine+) that focuses on target instruction coverage to determine interesting states.

We tested our subjects with \APPR and competing approaches using a test budget of one hour, which is 
a common choice in several App testing papers~\cite{Gu:APE:ICSE:2019, Borges-Droidmate2-ASE-2018, Pan:QTesting:2020, Guo:GESDA:2020}. 

We executed each tool with each updated version ten times. For \APPR, for each of these ten experiments, we simulated a realistic usage scenario by first testing the initial version of the App considering the entire code as updated, thus deriving an initial App model for V0. We then tested the upgraded versions by reusing the App model generated for the previous App version considered in the same experiment. 
In total, the experiment took 6940 hours of computing time.

In our experiments, we determine the statistical significance of the difference using a non-parametric Mann-Whitney U-test (with $\alpha$= 0.05). Further, we discuss effect size by relying on  the Vargha and Delaney’s $A_{12}$ statistics~\cite{VDA}, a non-parametric effect size measure.



\MAJORBEGIN{}
\subsection{RQ1 - Test oracle cost}
\subsubsection{Experimental Design}
\label{sec:rq0:design}

To address RQ1, as in previous work~\cite{ATUA}, we count the number of inputs generated by each testing tool, for each test execution
run. For \CALM and ATUA, we rely on the CSV file generated by the ActionTrace component of ATUA,
which reports all the inputs triggered during testing. For Monkey, we process the log to determine the number of inputs generated. For all the other tools, we record the number
of test inputs reported by the tool at the end of their execution.

For each subject App, we compare \emph{distributions of the number of inputs generated across tools}. 
However, since in CALM the number of selected outputs does not simply match the number of inputs but is further reduced by reporting only the UTAs, we report both the number of inputs generated by \CALM and the final number of UTAs selected by CALM.
To answer positively this research question, \APPR should not generate significantly more inputs than ATUA, and generate fewer test inputs than other tools. Further, the number of UTAs selected by \CALM should be significantly lower than the number of inputs generated by other approaches. This is the most important evaluation criterion since, with \CALM, engineers only inspect the output screens produced after UTAs.

\subsubsection{Results}

\begin{figure}[tb]
\includegraphics[width=12cm]{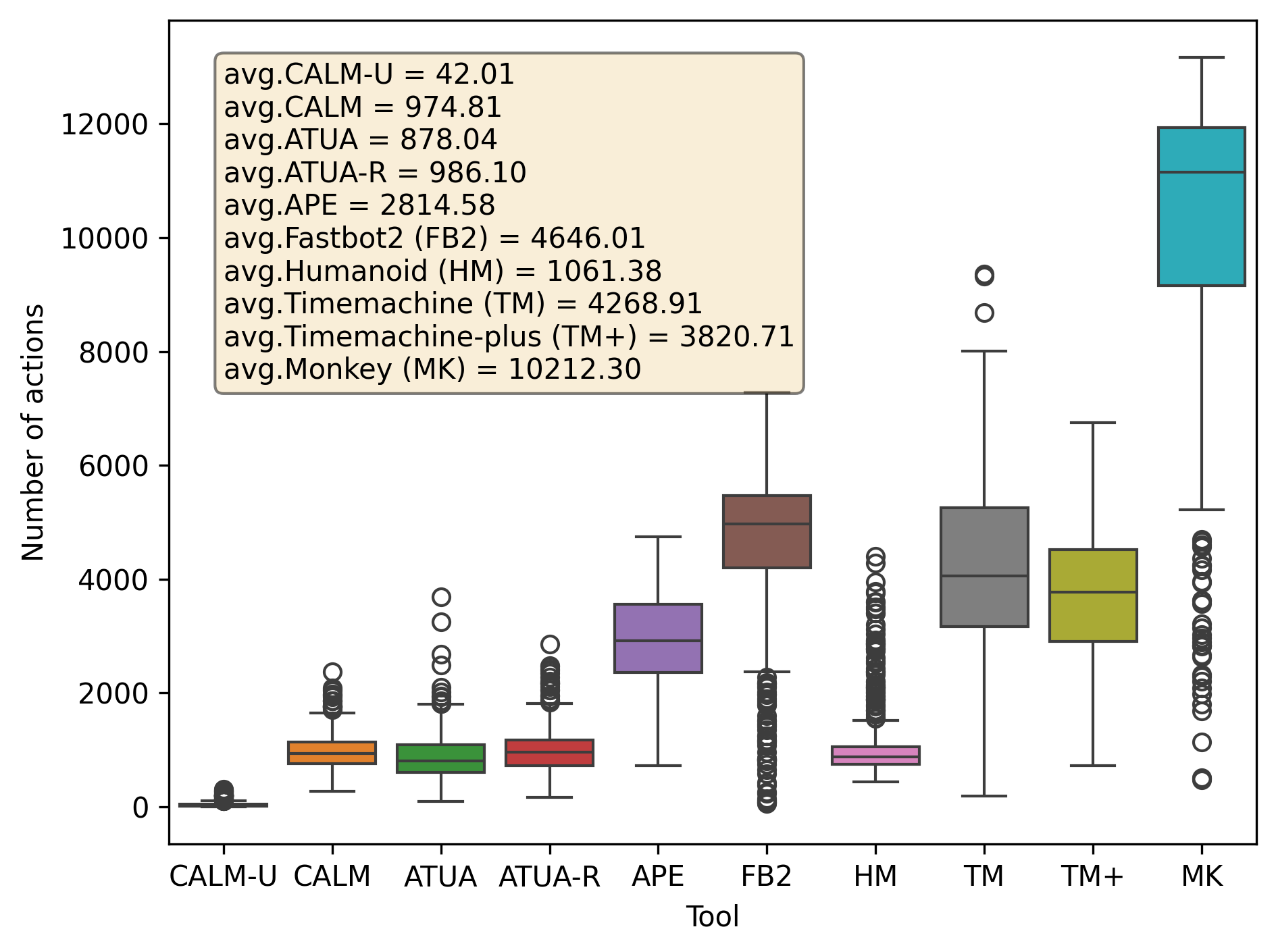}
\caption{Distribution of inputs generated by the different approaches considered in our experiments.}
\label{fig:inputs:dist}
\end{figure}
\begin{figure}[tb]
    \includegraphics[width=12cm]{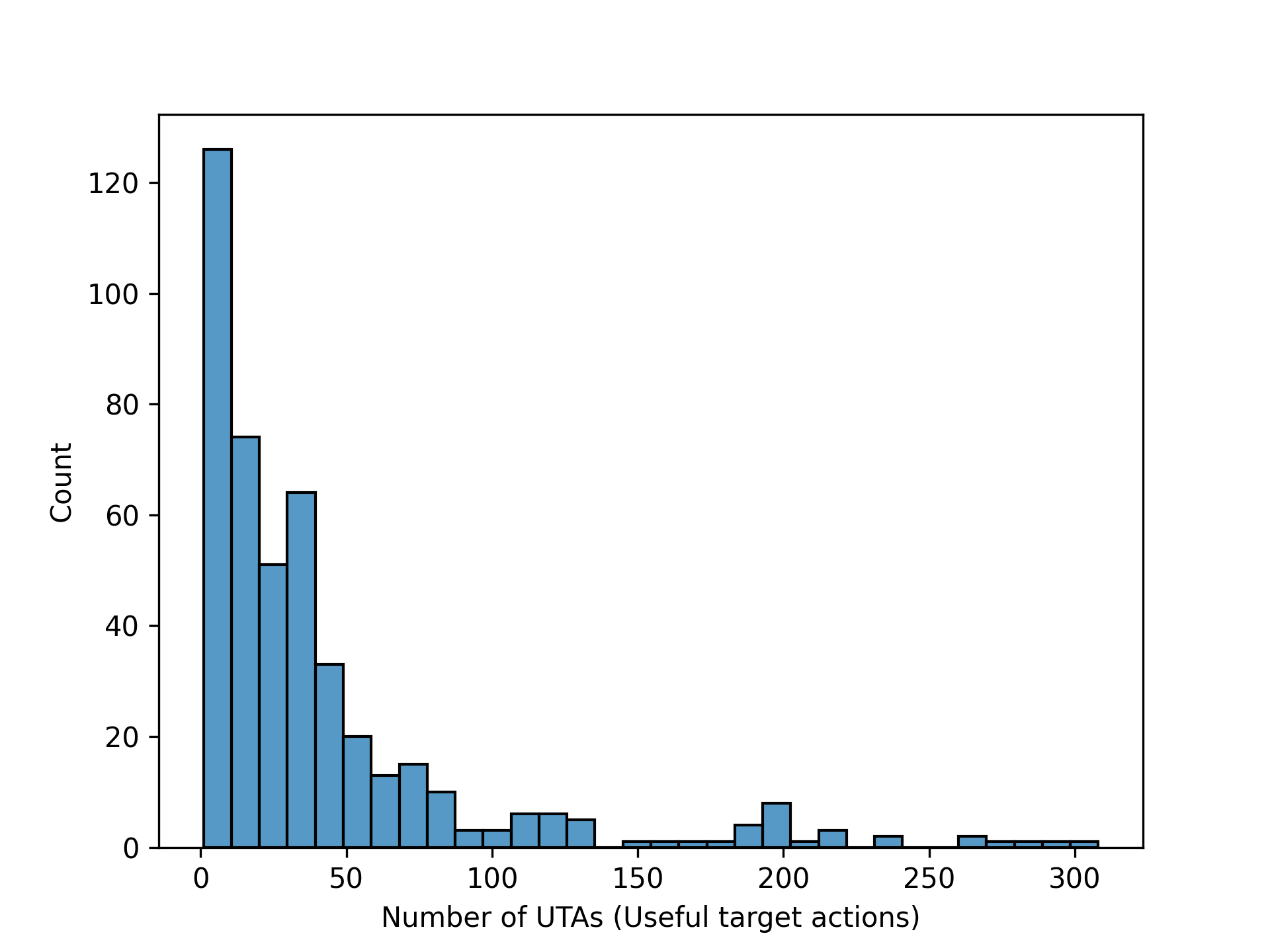}
    \caption{Distribution of UTAs (Useful Target Actions) generated by CALM}
    \label{fig:enter-label}
\end{figure}

\begin{table}[htb]
    \centering
    \caption{Statistical significance and effect size for number of inputs generated by the different approaches.}
  \scriptsize
    \begin{tabular}{|c|c|c|c|c|c|c|c|c|c|c|c|}
      
        \hline
        \multicolumn{12}{|c|}{\textbf{MannWhitney U-test’s P-value} }\\
        \hline
         &&(1)&(2)&(3)&(4)&(5)&(6)&(7)&(8)&(9)&(10)\\
        CALM-U & (1) & - & <0.05 & <0.05 & <0.05 & <0.05 & <0.05 & <0.05 & <0.05 & <0.05 & <0.05 \\
        CALM & (2) & <0.05 & - & <0.05 & 0.352 & <0.05 & <0.05 & 0.169 & <0.05 & <0.05 & <0.05 \\
        ATUA & (3) & <0.05 & <0.05 & - & <0.05 & <0.05 & <0.05 & <0.05 & <0.05 & <0.05 & <0.05 \\
        ATUA-Reuse & (4) & <0.05 & 0.352 & <0.05 & - & <0.05 & <0.05 & 0.752 & <0.05 & <0.05 & <0.05 \\
        APE & (5) & <0.05 & <0.05 & <0.05 & <0.05 & - & <0.05 & <0.05 & <0.05 & <0.05 & <0.05 \\
        Fastbot2 & (6) & <0.05 & <0.05 & <0.05 & <0.05 & <0.05 & - & <0.05 & <0.05 & <0.05 & <0.05 \\
        Humanoid & (7) & <0.05 & 0.169 & <0.05 & 0.752 & <0.05 & <0.05 & - & <0.05 & <0.05 & <0.05 \\
        TimeMachine & (8) & <0.05 & <0.05 & <0.05 & <0.05 & <0.05 & <0.05 & <0.05 & - & <0.05 & <0.05 \\
        TimeMachine+ & (9) & <0.05 & <0.05 & <0.05 & <0.05 & <0.05 & <0.05 & <0.05 & <0.05 & - & <0.05 \\
        Monkey & (10) & <0.05 & <0.05 & <0.05 & <0.05 & <0.05 & <0.05 & <0.05 & <0.05 & <0.05 & - \\
        
        \hline
        \multicolumn{12}{|c|}{\textbf{$A_{12}$ effect size} }\\
        \hline
        &&(1)&(2)&(3)&(4)&(5)&(6)&(7)&(8)&(9)&(10)\\
        CALM-U & (1) & - & $\approx$0 & $\approx$0 & $\approx$0 & 0 & 0.002 & 0 & $\approx$0 & 0 & 0 \\
        CALM & (2) & 1 & - & 0.6 & 0.517 & 0.02 & 0.037 & 0.525 & 0.019 & 0.01 & 0.007 \\
        ATUA & (3) & 1 & 0.4 & - & 0.421 & 0.02 & 0.035 & 0.413 & 0.018 & 0.009 & 0.006 \\
        ATUA-Reuse & (4) & 1 & 0.483 & 0.579 & - & 0.021 & 0.037 & 0.506 & 0.02 & 0.011 & 0.007 \\
        APE & (5) & 1 & 0.98 & 0.98 & 0.979 & - & 0.116 & 0.953 & 0.223 & 0.27 & 0.028 \\
        Fastbot2 & (6) & 0.998 & 0.963 & 0.965 & 0.963 & 0.884 & - & 0.959 & 0.624 & 0.715 & 0.062 \\
        Humanoid & (7) & 1 & 0.475 & 0.587 & 0.494 & 0.047 & 0.041 & - & 0.033 & 0.027 & 0.008 \\
        TimeMachine & (8) & 1 & 0.981 & 0.982 & 0.98 & 0.777 & 0.376 & 0.967 & - & 0.573 & 0.054 \\
        TimeMachine+ & (9) & 1 & 0.99 & 0.991 & 0.989 & 0.73 & 0.285 & 0.973 & 0.427 & - & 0.045 \\
        Monkey & (10) & 1 & 0.993 & 0.994 & 0.993 & 0.972 & 0.938 & 0.992 & 0.946 & 0.955 & - \\
        \hline
    \end{tabular}
    \label{table:number_of_inputs_siignificance}
\end{table}
Figure~\ref{fig:inputs:dist} shows boxplots of the distribution of the number of inputs generated by each approach across runs. With an average of 878.04 inputs exercised for each App version under test, ATUA is being confirmed as the approach that minimizes the number of inputs generated during testing. However, CALM is slightly worse, with 974.81 inputs on average. Further, CALM performs slightly better than ATUA-R, which generates 986.10 inputs on average. The only other approach that performs similarly to CALM and ATUA is Humanoid, with 1061.38 inputs. Further, in Figure~\ref{fig:inputs:dist}, CALM-U captures the number of UTAs selected by \CALM for inspection, with an average of 42.01, and clearly shows that focusing on UTAs helps significantly reduce the number of outputs to be inspected.

Table~\ref{table:number_of_inputs_siignificance} provides information about the significance of  differences and VDA score for each pair of approaches.  Note that, regarding VDA, the approach on the first column performs better than the approaches on the other columns if the VDA score is below 0.5; indeed, we compare approaches in terms of number of generated inputs since it is desirable to minimize such inputs.

Table~\ref{table:number_of_inputs_siignificance} shows that the differences between CALM and ATUA are significant but effect size is small (i.e., 0.6); therefore, although ATUA remains the approach minimizing the number of inputs being generated, \CALM fares similarly well. 
Further, Table~\ref{table:number_of_inputs_siignificance}, confirms that \CALM performs similarly to ATUA-R and Humanoid.

However, recall that CALM does not report all the App output screens for visual inspection, but only the ones observed after useful target actions (UTAs), which greatly reduces test oracle cost. Figure~\ref{fig:enter-label} shows the distribution of UTAs across our subject Apps;
they range between 1 and 308 but most App versions can be tested with less than 40 UTAs (i.e., it is sufficient to verify up to 40 output screens). 
CALM-U in Table~\ref{table:number_of_inputs_siignificance} shows that the number of outputs to be inspected with \CALM is significantly lower than the number of outputs required by the  other approaches in our study, including \ATUA.

In conclusion, our results show that \CALM is one of the approaches generating the lowest number of inputs, which, combined with the UTA selection strategy, enables \CALM to select a much smaller number of outputs for manual inspection.




\MAJOREND{}

\subsection{RQ2 - Effectiveness for a given test budget}

\subsubsection{Experimental Design}
\label{sec:rq1:design}
Since we are interested in exercising code that is likely affected by changes (updated methods), we compare \APPR with the other approaches in terms of percentage of covered updated methods (hereafter, target method coverage) and instructions belonging to updated methods (hereafter, target instruction coverage). 

Since the identification of functional faults can only be based on the visual inspection of App screens rendered after every input action, it is necessary to compare the coverage results obtained when a similar number of App screens is inspected, so that we can assume the effort required to detect faults is similar across competing approaches. 

We assume that engineers apply the strategy presented in Section~\ref{sec:approach}: they inspect only the App screens generated by useful target actions (UTAs), which are defined as actions contributing to increasing the coverage of instructions belonging to updated methods. In our analysis, we therefore compare the coverage obtained for a same number of UTAs, which enables comparing effectiveness for a same fault detection cost.

During testing, we identify UTAs and the target instructions they cover.
Then, for each subject version, we compute the average number $N$ of UTAs generated by CALM, and we select, for each execution of the other testing tools on the same subject, the first $N$  UTAs being triggered. 
We then compute the target method and instruction coverage achieved with the selected UTAs.
We extended ATUA, APE, and TimeMachine to collect the instructions covered by each UTA. 
For Monkey and Fastbot2, it was not possible to implement the same extension; therefore, we report the coverage achieved by these tools with all the inputs triggered in one hour. Additionally, for completeness, we report the target coverage achieved by APE and ATUA with all the inputs triggered in one hour. Please note that, in practice, it would be infeasible for engineers to visually inspect all the App screens rendered with Monkey, APE, and Fastbot2 because of the large number of inputs they trigger~\cite{ATUA}; however, Monkey enables us to gain insights about the input space (i.e., how simple it is to exercise target methods without guidance).

To positively answer RQ2, \CALM should achieve a significantly higher target method and instruction coverage than competing approaches, for a same number of UTAs. 
Further, since performance fluctuations across App versions might be expected, we report on the number of versions in which \CALM performs better.
To this end, we rely on the Vargha and Delaney’s $A_{12}$ statistics~\cite{VDA} 
applied to the ten execution results obtained for a given version.
Following standard practice, \CALM is deemed to perform better than other approaches when the difference is statistically significant and $A_{12} > 0.56$.

\subsubsection{Results}


\begin{figure*}
    \begin{subfigure}[b]{9cm}
        \includegraphics[width=9cm]{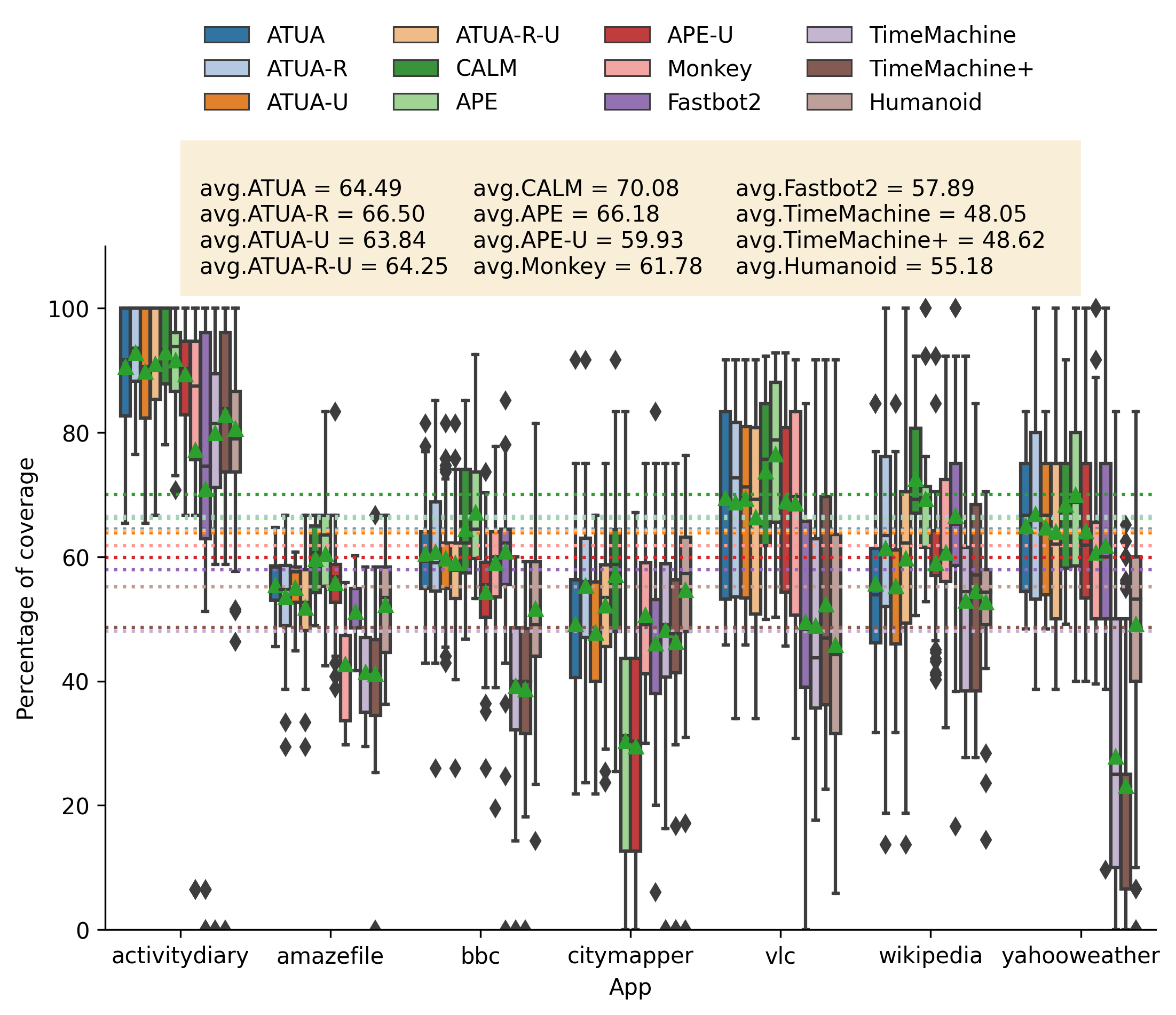}
        \caption{
            Target method coverage.}
        \label{fig:RQ1:updatedMethod}
    \end{subfigure}
    
    \vspace{1em}
    \begin{subfigure}[b]{9cm}
        \includegraphics[width=9cm]{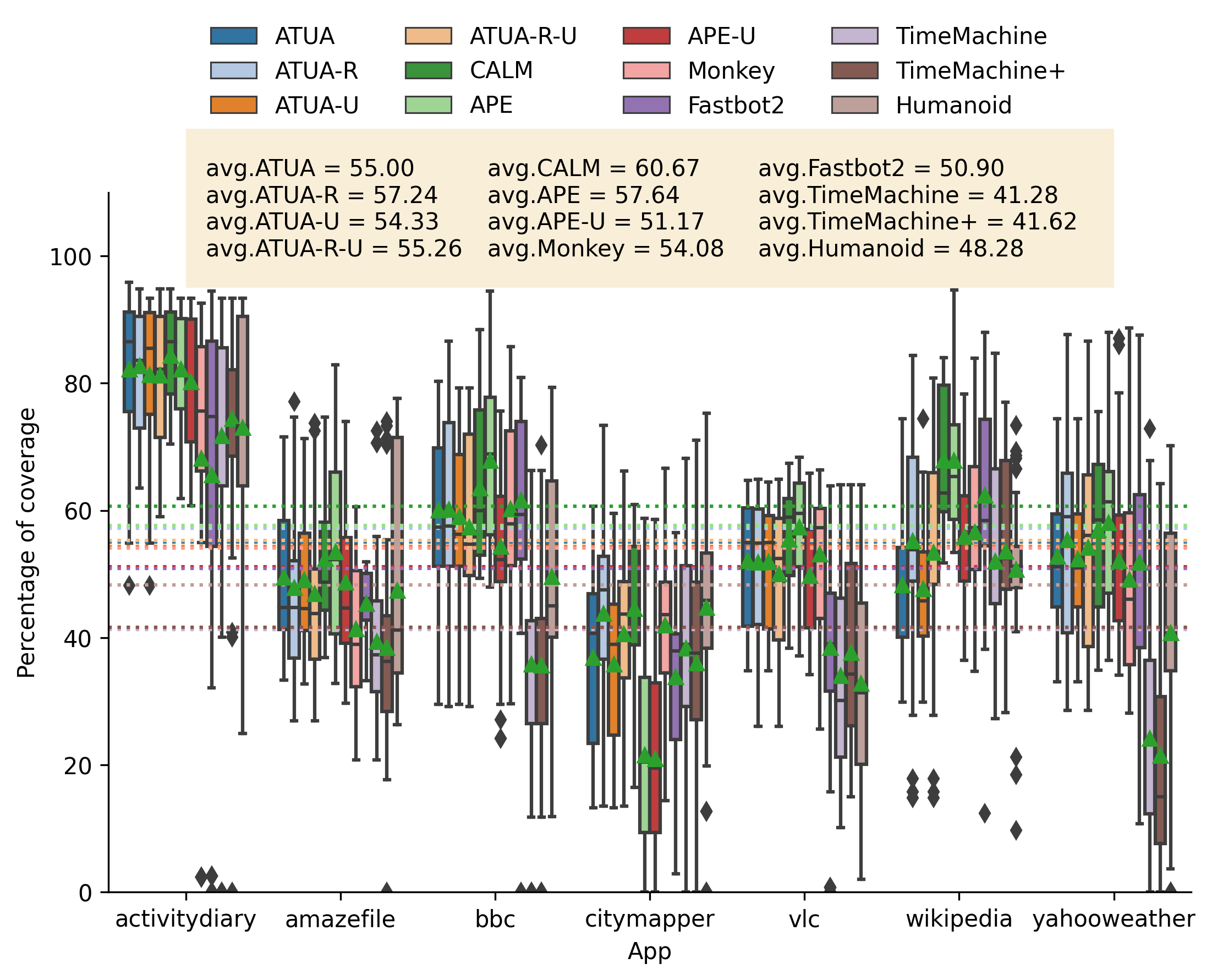}
        \caption{
            Target instruction coverage.}
        \label{fig:RQ1:updatedinstruction}
    \end{subfigure}
    \caption{Distribution of target method coverage and target instruction coverage. }
    \label{fig:RQ1}
\end{figure*}

\begin{table}[tb]
\footnotesize
\caption{Number of versions in which \APPR performs significantly better than competing approaches and vice-versa. }
\label{table:RQ1_versions}
\begin{tabular}{|
@{\hspace{0pt}}>{\raggedleft\arraybackslash}p{20mm}@{\hspace{1pt}}|
@{\hspace{0pt}}>{\raggedleft\arraybackslash}p{15mm}@{\hspace{1pt}}|
@{\hspace{0pt}}>{\raggedleft\arraybackslash}p{15mm}@{\hspace{1pt}}|
@{\hspace{0pt}}>{\raggedleft\arraybackslash}p{15mm}@{\hspace{1pt}}|
@{\hspace{0pt}}>{\raggedleft\arraybackslash}p{15mm}@{\hspace{1pt}}|
}
\hline
\multirow{2}{*}{\textbf{Tool}}   & \multicolumn{2}{c|}{\textbf{Target method coverage}} & \multicolumn{2}{c|}{\textbf{Target instr. coverage}} \\
\cline{2-5}
    & \multicolumn{1}{c|}{\CALM better}           & \multicolumn{1}{c|}{\CALM  worse}       & \multicolumn{1}{c|}{\CALM better}             & \multicolumn{1}{c|}{\CALM worse}             \\ 
\hline 
ATUA-U              & 41 (78.85\%)              & 0 (0\%)               & 46 (88.46\%)                & 0 (0\%)                  \\ \hline 
ATUA-R-U              & 37 (71.15\%)              & 2 (3.85\%)               & 37 (71.15\%)                & 2 (3.85\%)                  \\ \hline 
APE-U & 44 (84.62\%)              & 2 (3.85\%)               & 50 (96.15\%)                & 0 (0\%)                 \\ \hline 
Monkey                      & 35 (67.31\%)              & 4 (7.69\%)               & 35 (67.31\%)                & 4 (7.69\%)  \\  \hline             
Fastbot2                      & 38 (73.08\%)              & 1 (1.92\%)               & 37 (71.15\%)                & 2 (3.85\%)  \\  \hline   
TimeMachine                   & 48 (92.31\%)               &  0 (0\%)            & 46 (88.46\%)               & 0 (0\%) \\  \hline 
TimeMachine+                   & 44 (84.62\%)               &  1 (1.92\%)            & 42 (80.77\%)               & 1 (1.92\%) \\  \hline 
Humanoid                   &  40 (76.92\%)               &  1 (1.92\%)            &  38 (73.08\%)               & 1 (1.92\%) \\  \hline
\end{tabular}
\end{table}


Figures~\ref{fig:RQ1:updatedMethod} and~\ref{fig:RQ1:updatedinstruction} show the distributions of the target method and instruction coverage for \APPR, for each subject App. 
The two Figures show similar distributions; 
a data point is the coverage achieved with one test execution on one App version. \textbf{\APPR is the approach yielding the best results, on average, with 70.08\% and 60.67\% target method and instruction coverage}, respectively; differences between CALM and other tools are statistically significant. The second-best result is obtained by APE (66.18\% and 57.64\%), if we consider all the inputs generated. However, to be realistic, we should rely exclusively on UTAs, which make the performance of APE (i.e., APE-U) drop to 59.93\% and 51.17\%, approximately a 10\% and 9\% decrease from CALM, respectively. APE performs better than Fastbot2 (57.89\% and 50.90\%), which differs from previous results~\cite{Fastbot2}, likely because Fastbot2 overfits the specific industrial scenarios for which it was developed.

ATUA-U performs better than APE-U (63.84\% and 54.33\%), while ATUA-R-U performs slightly better than ATUA-U (64.25\% and 55.26\%) but differences are not significant, which indicates that model reuse alone provides limited benefits without all the heuristics integrated into \CALM (\CALM performs significantly better than ATUA-R-U).

\CALM performs significantly better than APE-U, ATUA-U, and ATUA-R-U thus showing that model reuse improves the testing of updated Apps but \CALM's heuristics are necessary to effectively reuse models (indeed, \CALM performs significantly better than ATUA-R-U). 
Further, the better performance of \CALM over Fastbot2 shows that \textbf{model reuse alone, without appropriate strategies to drive testing, is not sufficient to effectively test updated methods}. 

The need for appropriate testing strategies is also highlighted by the poor performance of Monkey, TimeMachine, and Humanoid. The performance of Humanoid and TimeMachine does not change when either considering all the inputs or UTAs only; for such reason we do not report Humanoid-U and TimeMachine-U in Figure~\ref{fig:RQ1}. TimeMachine is likely negatively affected by the cost of taking execution snapshots. TimeMachine+ is the second-worst approach, thus showing that \textbf{focusing on target instructions is not sufficient to test modified functionalities} but a dedicated approach (i.e., \CALM) is needed. Humanoid poorly performs likely because it tends to focus on the main App features, which might not be the modified ones, in addition to being affected by other limitations~\cite{MuBot}. 
Please note that both Fastbot2 and Monkey also 
lead to a much higher number of output screens to be manually verified (i.e., 4645, for Fastbot2, and 51,500, for Monkey, on average for each version versus a range between 1 and 186 for \CALM, 35 on average). 


Table~\ref{table:RQ1_versions} provides the number of App versions in which \ATUA performs significantly better than competing approaches and vice-versa.
Table~\ref{table:RQ1_versions} shows that \textbf{\CALM performs significantly better than competing approaches for a significantly larger number of versions}, thus showing it is the best choice to incrementally test App versions.

\subsection{RQ3 - Effectiveness over time}

\subsubsection{Metrics}


We study the effectiveness of \APPR, for increasing testing time budgets and updates of different magnitude. To measure such magnitude we rely on the proportion of updated methods because it enables us to compare results achieved with Apps of different sizes. 
 
 Based on the distribution of the number of App versions per percentage of updated methods in our subjects, we identified three distinct patterns in App development (e.g., from bug fixes to major releases). Tiny updates with [0\%,1\%) updated App methods are very frequent (52.85\% of our versions); small updates with [1\%,10\%) updated methods are relatively frequent (34.62\% of versions); medium updates with [10\%,30\%) updated methods are much less frequent (11.54\% of versions). 

Like in RQ1, we rely on code coverage as a proxy for effectiveness. We focus on target instruction coverage because method coverage is likely to show high variations between test executions when updates have limited magnitude.

During RQ2 experiments, we traced timestamps and the target instruction coverage for every input action. To address RQ3, we focus on the coverage achieved by each technique, after every minute, considering  UTAs only, as in RQ2. 

In our analysis, we exclude Monkey and Fastbot2 since 
they are not practically applicable in our context given that that they do not enable the selection of UTAs.

For each update size range, we compute the average target instruction coverage for all the ten experiment runs of all the App versions having a number of updated methods in that range; we discuss the significance of their difference across ranges based on the Mann Whitney U-test (with $\alpha= 0.05$).

\subsubsection{Results}

\begin{figure*}
    \begin{subfigure}[b]{8cm}
        \includegraphics[width=8cm]{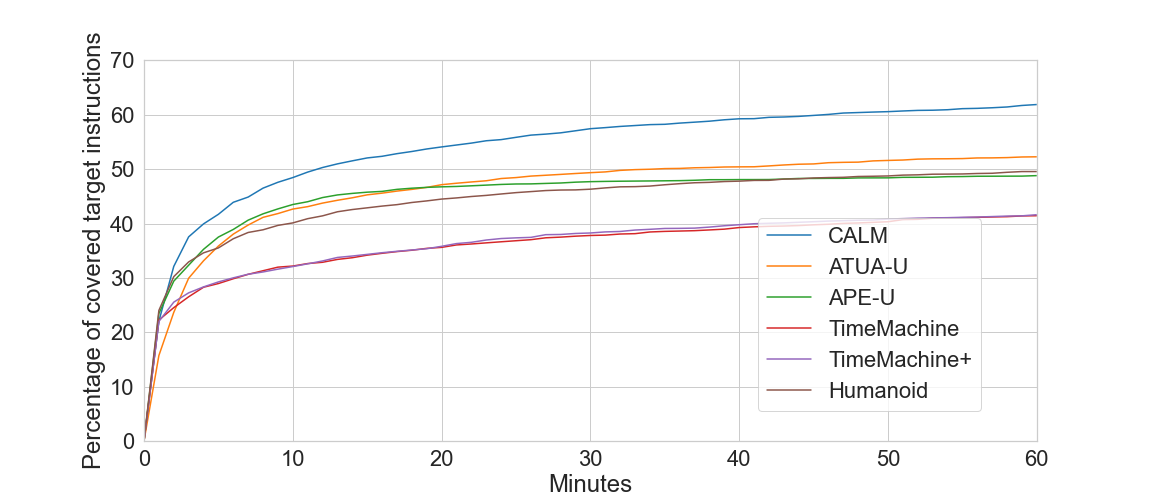}
        
        \caption{Up to 1\%}
    \end{subfigure}
    \begin{subfigure}[b]{8cm}
        \centering
        \includegraphics[width=8cm]{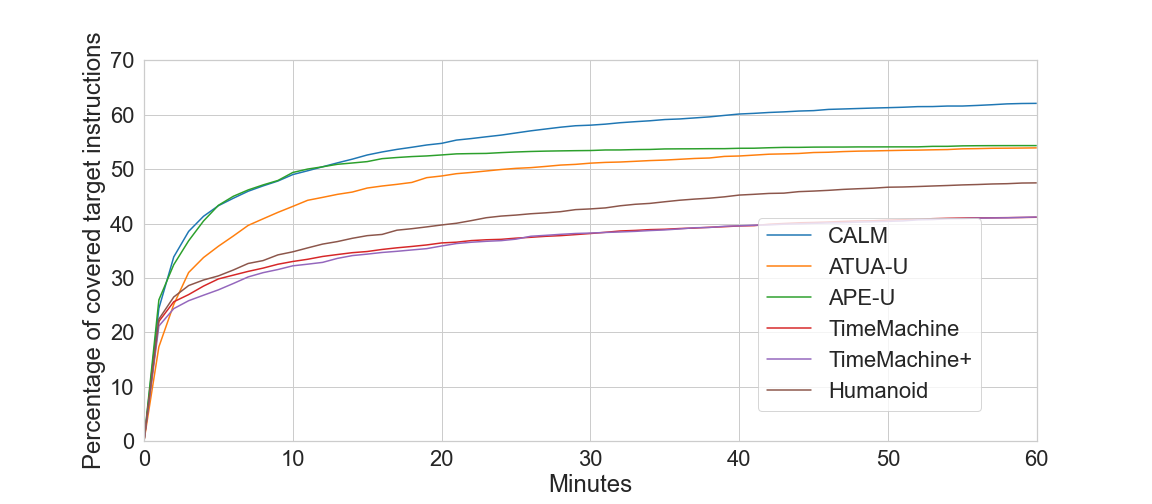}     
        \caption{Between 1\% and 10\%}
    \end{subfigure}
    \begin{subfigure}[b]{8cm}
        \centering
        \includegraphics[width=8cm]{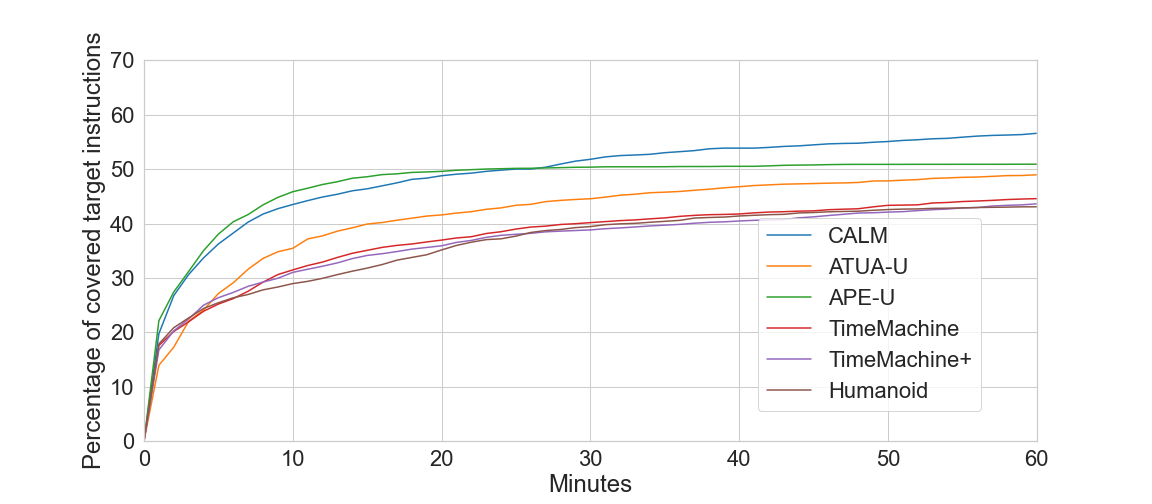}  
        \caption{More than 10\%}
    \end{subfigure}
       \caption{
      Average \% of covered target instructions across subjects, grouped by magnitude of changes (updated App methods).}
            \label{fig:RQ2:incrementalTargetInstructionCoverage}
\end{figure*}

Figure~\ref{fig:RQ2:incrementalTargetInstructionCoverage} depicts the average target instruction coverage over time.
Our results show that \textbf{\CALM always achieves higher average target instruction coverage than \ATUA, for any test budget}, which indicates that 
model reuse, including all the \CALM's optimizations, is always the best choice; this would not have been the case if the reused models were driving \CALM towards exercising obsolete input sequences leading to unexpected App states.
\CALM always performs significantly better than Humanoid, TimeMachine, and TimeMachine+.
APE-U, however, performs slightly better than \CALM in the first minutes of execution (e.g., model loading cost may negatively affect \CALM), but then \CALM overcomes APE-U. Interestingly, the difference in performance between the two approaches and the moment in which \CALM takes over depends on the magnitude of the change.

With up to 1\% updated methods, which is the most frequent case (more than half of our subject versions), 
APE-U performs significantly better only in the first minute of execution but with a limited improvement of 1.4 percentage points (pp).
CALM starts faring significantly better than APE-U after 2 minutes of execution with the average difference between CALM and APE-U increasing from 5 (at 2 minutes of testing) to 13 pp (after 60 minutes).

With 1\% to 10\% updated App methods, APE-U performs significantly better only in the first minute (1.6 pp higher). Between 15 and 32 minutes the difference between the two is not significant but \CALM's coverage increases linearly from 1.2 pp to 4.75 pp. After 32 minutes the difference is significant (5 pp higher) and then reaches 7.5 pp at 60 minutes.
\CALM reaches a max average coverage of 62.1\% versus 54.3\% for APE-U; further, APE-U reaches a plateau  at 30 minutes (in the last 30 minutes of APE execution, its mean coverage increases only by 0.9 pp), while \CALM keeps improving its coverage.

When the proportion of updated methods is large (more than 10\%), APE-U performs  slightly better than \CALM in the first 26 minutes but differences are not significant and the improvement is moderate (up to 2.4 pp); however, APE-U reaches a plateau at 23 minutes while \CALM keeps improving. After 38 minutes the difference between the two approaches is significant, with \CALM performing better by 5.6 pp after 60 minutes. 
When the magnitude of changes is large, a larger test budget is required to observe a significant difference between \CALM and APE-U. Such result is expected since, with a larger proportion of updated methods, it is easier to exercise updated methods regardless of the guidance effectiveness. However, the difference between \CALM and APE-U keeps increasing for a larger test budget.

\begin{figure*}
        \includegraphics[width=12cm]{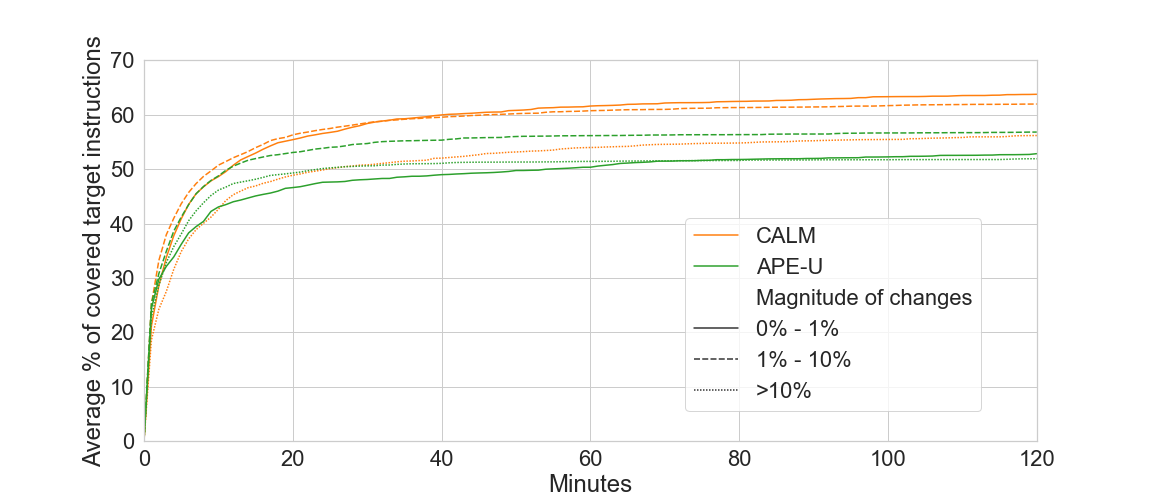}
       \caption{
      Average \% of covered target instructions across subjects up to 2 hours, grouped by magnitude of changes (updated App methods).}
            \label{fig:RQ2:incrementalTargetInstructionCoverage_2h}
\end{figure*}

We performed an additional experiment with \CALM and APE-U executed for two hours. Results are shown in Figure~\ref{fig:RQ2:incrementalTargetInstructionCoverage_2h}\footnote{Please note that the experiment for a 2-hour budget corresponds to new executions of \CALM and APE on all the subject Apps; therefore, since the datapoints are not the same as the ones collected for the 1-hour budget, the curves in Figures~\ref{fig:RQ2:incrementalTargetInstructionCoverage} and~\ref{fig:RQ2:incrementalTargetInstructionCoverage_2h} do not exactly match but show similar trends.}; when more than 10\% of App methods are updated, after two hours, \CALM and APE-U achieve a target instruction coverage of 56.19\% and 51.92\%, respectively. The difference between the two approaches is significant and increases from 2.56 pp (1-hour budget) to 4.26 pp (2-hour budget).

To summarize, \CALM always performs \textbf{significantly better than Humanoid, TimeMachine, and TimeMachine+}. Also, \textbf{for tiny updates (the majority) \CALM performs better than APE-U after 2 minutes of test budget, which is reasonable. For larger updates, the larger budget required by \CALM is justified by its coverage not reaching a plateau but steadily improving 
until becoming significantly higher than that of APE-U.}

\MAJORBEGIN{}
\MAJOR{R2.5}{\subsection{RQ4 - Functional Faults Detection Capability}}\label{rq4}
\subsubsection{Experimental Design}
To study \CALM's fault detection capabilities, we consider real functional faults affecting our subjects Apps. 
To perform our study, we need to determine if the fault had been exercised by a testing tool (e.g., by determining if the source code location of the fault is covered) and if an App output shows a failure. Therefore, we focus on open-source Apps, which come with source code and an issue tracker system not available for proprietary Apps. Among our subject Apps, the open-source Apps with faults in their issue tracking systems include ActivityDiary~\cite{activitydiary} and AmazeFileManager~\cite{amazefile}.

For ActivityDiary and AmazeFileManager, consistent with the objectives of \CALM, we identified all the bug reports concerning functional faults leading to failures  by inspecting the App screen. Among these reports, we selected the ones concerning faults that had been fixed because the availability of a bug fix simplifies fault understanding (e.g., we can look at the patches to determine the faulty lines of code). Finally, we selected the faults that we could reproduce by manually executing the faulty App version, thus ending up with 11 faults. Although 11 faults might not be sufficient to report on the significance of the differences between CALM and APE-U, we expect the analysis to provide qualitative insights.

Since not all these 11 faults had been introduced in the App versions selected for our study, we created additional faulty versions for the selected Apps by introducing faulty updates, as follows:

\begin{itemize}
    \item If the faulty method is among the target methods for an App version previously tested with CALM, we select such version as the target for our experiment.
    \item If an App version tested with CALM includes the faulty method but such method is not among the target methods for that App version (i.e., the fault was introduced previously), we force CALM to test such faulty method by introducing a change in it (i.e., we introduce a new logging instruction).
    \item If the faulty method does not belong to any App version tested with CALM, we create a target App version by reintroducing the fault into one of the selected App versions.
\end{itemize}

\renewcommand{\arraystretch}{1.5}
\begin{table}[]
    \caption{Functional bugs considered in the preliminary evaluation}
    
    \label{table:bugs}
    \footnotesize
    
    \centering
     \begin{tabular}{|p{0.8cm}|p{0.6cm}|p{5cm}|p{0.8cm}|p{0.8cm}|>{\centering\arraybackslash}p{0.8cm}|p{0.8cm}|p{1.2cm}|}
        \hline
        \multicolumn{1}{|>{\centering\arraybackslash}p{0.8cm}|}{FaultId}  & \multicolumn{1}{>{\centering\arraybackslash}p{0.6cm}|}{App} & \multicolumn{1}{>{\centering\arraybackslash}p{5cm}|}{Fault description} & \multicolumn{1}{>{\centering\arraybackslash}p{0.8cm}|}{Github issue ID} & \multicolumn{1}{>{\centering\arraybackslash}p{0.8cm}|}{Test version} & \multicolumn{1}{>{\centering\arraybackslash}p{0.8cm}|}{Is reintroduced} & \multicolumn{1}{>{\centering\arraybackslash}p{0.8cm}|}{Number of target methods} & \multicolumn{1}{>{\centering\arraybackslash}p{1.2cm}|}{Failure type} \\
         \hline
        AC01 & AD & When the activity edit page is filled, if the screen is rotated then the filled content is lost.& \#53 & v134 & \checkmark & 49 & Display \\
        AC02 & AD & After clicking on the "Delete" button on an activity's edit page, the activity is still present in other pages. & \#59 & v111 & \checkmark & 18 & Interaction\\
        AC03 & AD & When clicking on a picture of an activity, nothing happens while it is supposed to open the picture with an image viewer. & \#162 & v134 & \checkmark &49 & Interaction\\
        AC04 & AD & When undoing the recent activation of an activity, the state of the activated activity is incorrect if there is no activated activity before the undone one.  & \#286 & v134 & & 49 & Interaction\\
        AM01 & AM & After manually selecting all file items, if the options menu is opened, the "Deselect All" menu item is not present. & \#996 & v3.4.1 & \checkmark & 651 & Display\\
        AM02 & AM & Triggering the "Select All" menu item when all file items are selected causes all items to be deselected. & \#953 & v3.4.1 & \checkmark & 651 & Interaction\\
        AM03 & AM & When entering a file/folder name starting with a dot, the dialogue does not warn the user that the file/folder will be hidden. & \#1235 & v3.4.1 & \checkmark & 651 & Display\\
        AM04 & AM & The preselected configuration dialogue for the App's color allow multiple choices with radio buttons instead of single-choice. & \#1044 & v3.4.1 & \checkmark & 651 & Interaction\\
        AM05 & AM & When creating a new file, a  file name ending with ".txt" is not allowed. & \#1231 & v3.4.1 & \checkmark  & 651 & Display\\
        AM06 & AM & When searching files, hidden files with matching patterns are shown too. & \#1467 & v3.4.1 &  & 651 & Display\\
        AM07 & AM & When closing the "Hidden Files" dialogue, the file list is not refreshed and, therefore, the updates from the dialogue do not appear. & \#1712 & v3.4.1 &  & 651 & Interaction\\
         \hline
    \end{tabular}
    \footnotesize
    
    \textbf{Legends.} \textbf{AD}: Activity Diary(https://github.com/ramack/ActivityDiary), \textbf{AM}: Amaze File manager (https://github.com/TeamAmaze/AmazeFileManager)
    
\end{table}
\renewcommand{\arraystretch}{1}

Table~\ref{table:bugs} shows the details of the 11 faults selected for our study; we also indicate if the fault was already present in the selected App version or if it has been reintroduced. At a high level, five faults lead to a failure that consists of a UI display issue (e.g.,  displaying incorrect information), and six faults lead to a UI interaction problem (e.g., the App does not produce any output after a user input). To summarize, for our study, we considered a copy of v3.4.1 of Amaze File Manager with 7 faults, a copy of 
v134 of Activity Diary with 3 faults, and a copy of Activity Diary v111 with one fault.

Since APE-U is the second-best approach based on the target instruction coverage studied in RQ2 and RQ3, we also compare \APPR with APE-U. 

We tested the subject Apps with \APPR and APE-U and visually inspected the screenshots taken after exercising a UTA to determine the presence of failures. For each execution, as per RQ2, we allocate a test budget of one hour. To deal with randomness, we tested each faulty version 10 times. With \CALM, for each App version, we start \CALM using the App model derived when testing the previous version of the App under test, derived for RQ2. 

For each fault, we compute the \emph{Fault Detection Probability (FDP)} as the proportion of test executions leading to at least one failure (i.e., the screenshot recorded after a UTA includes an erroneous output caused by one of the faults). 
Furthermore, to better investigate why failures are not observed in some cases, we report the proportion of executions in which the faulty methods have been exercised (hereafter, \emph{Faulty Method Coverage} --- FMC).

\subsubsection{Results}
\label{sec:rq4:results}
\begin{table}[tb]
\footnotesize
\caption{Fault coverage by \APPR and APE-U: we report Fault method coverage (FMC) and Fault Detection Probability (FDP) }
\label{table:RQ4}
\begin{tabular}{|
>{\centering\arraybackslash}p{12mm}|
>{\raggedleft\arraybackslash}p{12mm}|
>{\raggedleft\arraybackslash}p{12mm}|
>{\raggedleft\arraybackslash}p{12mm}|
>{\raggedleft\arraybackslash}p{12mm}|
}
\hline
\multicolumn{1}{|c|}{\multirow{2}{*}{\textbf{Fault Id}}}  & \multicolumn{2}{c|}{\textbf{\APPR}} & \multicolumn{2}{c|}{\textbf{APE-U}} \\ \cline{2-5}
    & \multicolumn{1}{c|}{FMC (\%)}           & \multicolumn{1}{c|}{FDP (\%)}       & \multicolumn{1}{c|}{FMC (\%)}             & \multicolumn{1}{c|}{FDP (\%)}             \\ 
\hline 
AC01&60&0&100&0 \\
AC02&100&70&100&90 \\
AC03&90&60&0&0 \\
AC04&100&50&100&40 \\
AM01&100&0&100&10 \\
AM02&100&40&100&20 \\
AM03&100&90&100&100 \\
AM04&100&0&0&0 \\
AM05&100&50&100&100 \\
AM06&100&10&30&0 \\
AM07&100&0&100&0 \\
\hline
\textbf{Average}&\textbf{95.45}&\textbf{33.64}&\textbf{75.45}&\textbf{32.73} \\
\hline
\end{tabular}
\end{table}
Table~\ref{table:RQ4} shows the FMC and FDP achieved by \APPR and APE-U. \APPR outperforms APE-U in terms of FMC as it achieves an average FMC of 95.5\%, compared to 75.5\% for APE-U. In two cases (i.e., AC03 and AM04) \APPR exercises faulty methods that are never reached by APE-U, thus highlighting the effectiveness of \CALM in reaching faulty methods.
\CALM performs slightly better than APE-U also in terms of FDP, with 33.64\% vs. 32.73\%.

Among the selected faults, five (i.e., AC02, AC04, AM02, AM03, and AM05) are detected by both approaches in at least one of the ten executions. 
Instead, two faults are detected only by \APPR and one only by APE-U. Overall, with ten test runs, \CALM detects 7 faults, while APE-U detects only 6. 

Our results suggest that no single approach detects a large proportion of faults if it is executed only once, for one hour. App testing tools should be executed multiple times; however, in practice, engineers do not have time to inspect all the App screens resulting from a large set of executions (recall from RQ1 that \CALM reports 40 App screens, on average, for each test execution). 
If we compare \CALM and APE-U based on the number of faults identified in two test runs, thus entailing a reasonable effort, we can notice that \CALM is likely to detect 5 faults (i.e., the ones with  $FDP >= 50\%$) while APE-U would detect 3, on average. Concluding, \CALM appears to be more effective than APE-U regarding fault detection, from a practical standpoint, though these results need to be confirmed by larger studies with more functional faults. Nevertheless, these results are consistent with those obtained based on coverage.



\MAJOREND{}

\MAJORBEGIN{}
\MAJOR{R2.4, R3.18}{\subsection{Discussion}}\label{discussion}

\begin{figure}
    \centering
    \begin{subfigure}[t]{0.45\textwidth}
        
        \includegraphics[width=\textwidth]{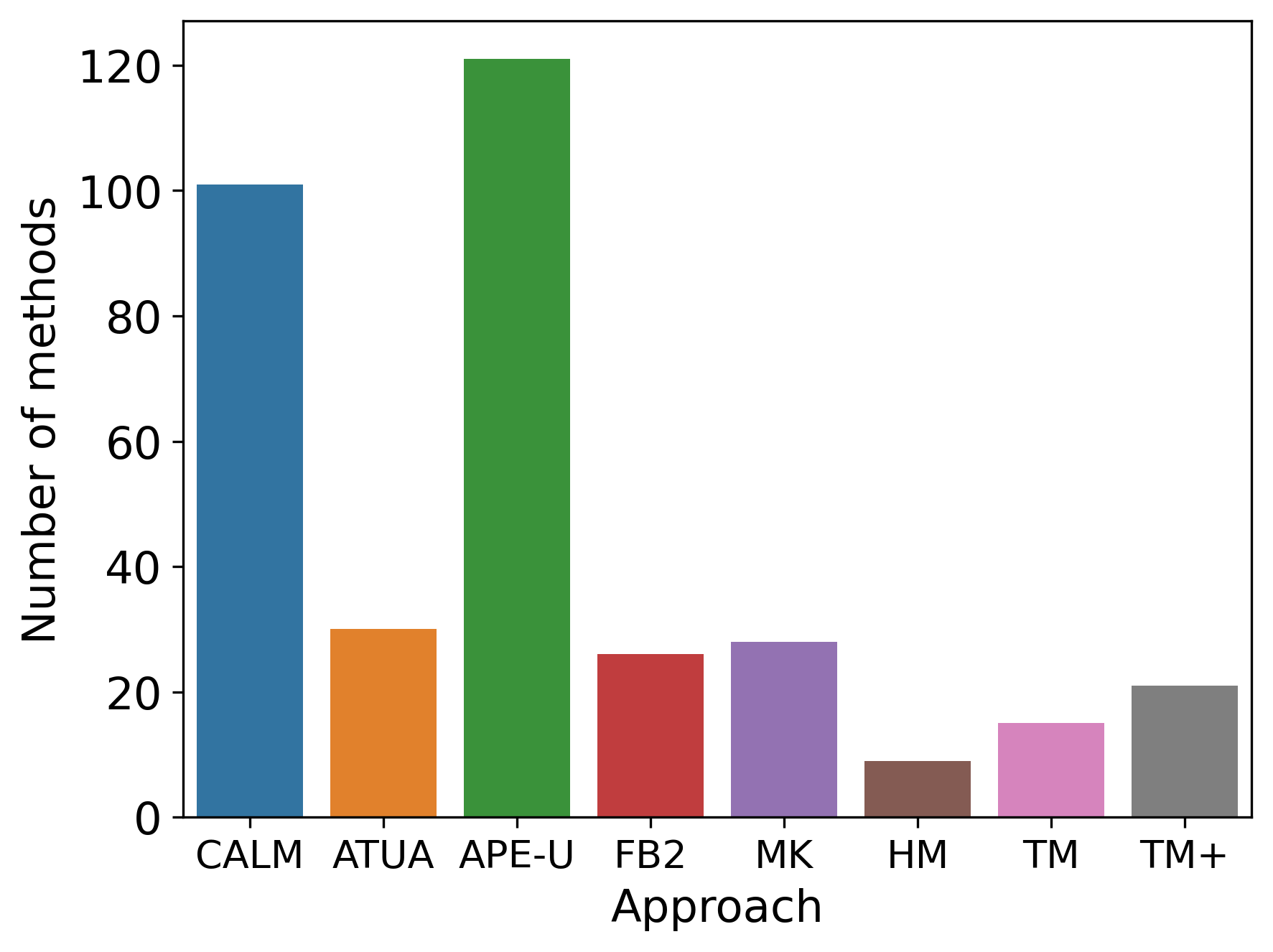}
        \subcaption{Number of target methods covered only by one approach.}
        \label{fig:target_methods_covered_diff:univocally_covered}
    \end{subfigure}
    
    
    \begin{subfigure}[t]{0.45\textwidth}
        \includegraphics[width=\textwidth]{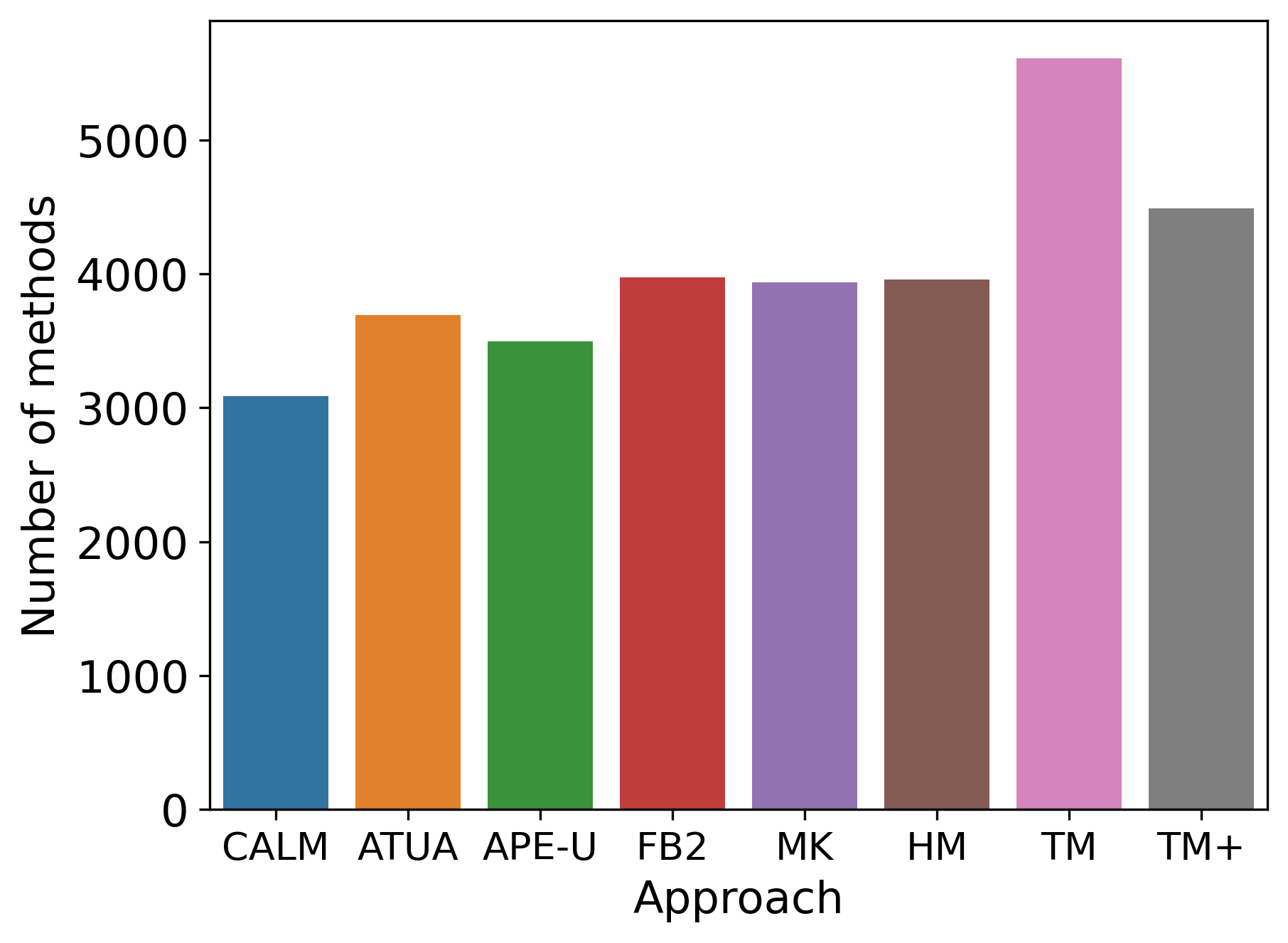}
        \subcaption{Number of target methods not covered by each approach.}
        \label{fig:target_methods_covered_diff:not_covered_by_each}
    \end{subfigure}
    \hfill
    \begin{subfigure}[t]{0.45\textwidth}
        \includegraphics[width=\textwidth]{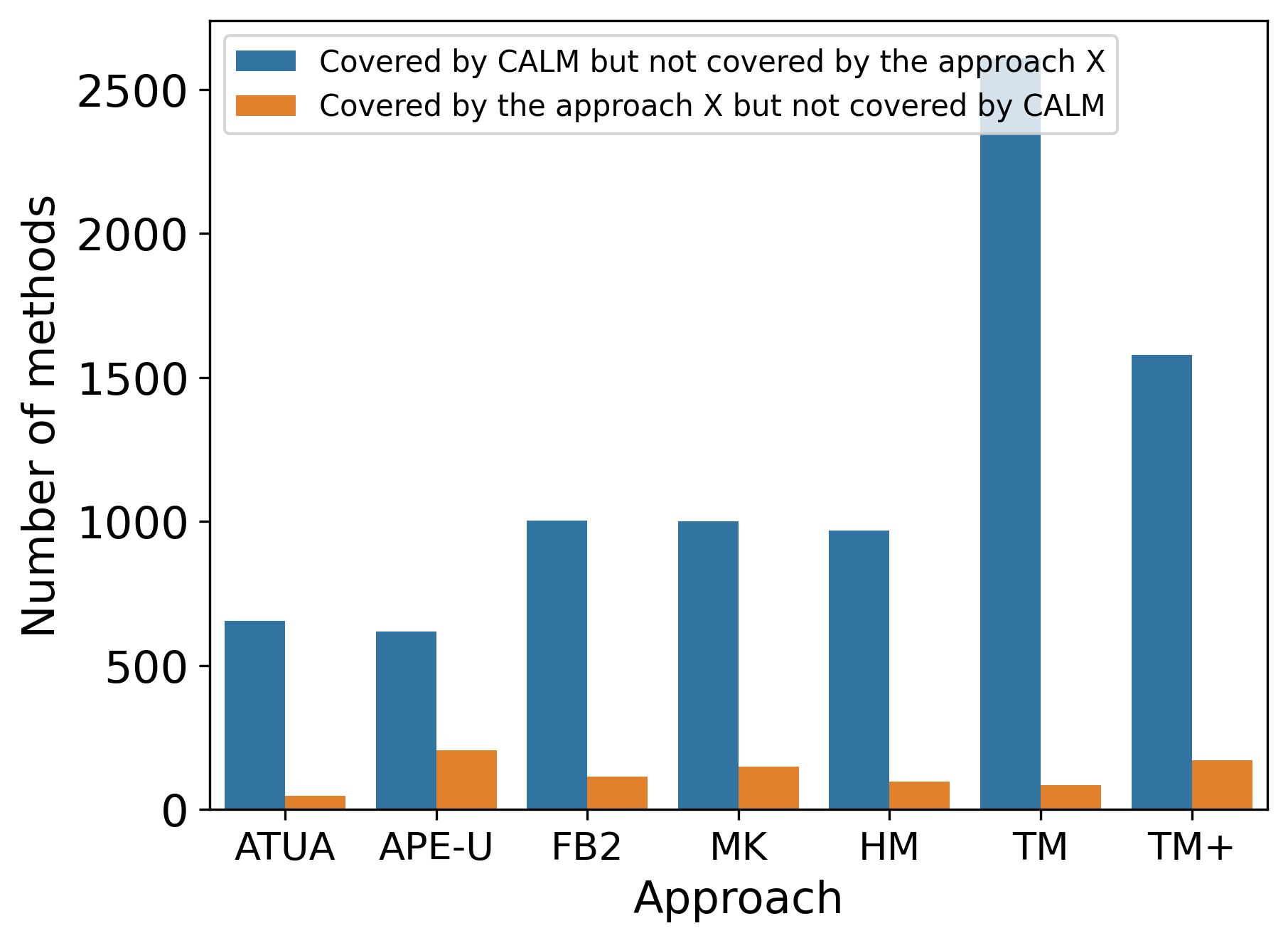}
        \subcaption{Number of target methods not covered by CALM but covered by a competing approach and vice versa.}
        \label{fig:target_methods_covered_diff:not_covered_by_CALM_but_another_and_inverse}
    \end{subfigure}

    \caption{Differences in target methods covered by CALM and other competing tools.}
    \label{fig:target_methods_covered_diff}
\end{figure}
Our results show that \CALM is the most effective approach to test App updates. 
However, our results also show that testing Apps remains an open problem; indeed, 
across all the tested versions, 2616 methods (25\%) are not exercised by any approach; further, only 33
\% of the faults can be detected by the two best approaches identified in our study (i.e., \CALM and APE-U). Below, we discuss the reasons that limit code coverage and fault detection capabilities of \CALM and competing approaches. 

\subsubsection{Limited code coverage}

We discuss the reasons why certain methods are not covered by any approach, and the degree of complementarity between \CALM and other approaches. 

Three are the reasons why 25\% of the methods are not covered by any approach:

\begin{itemize}
\item{\emph{Specific environment setup needed.}} 
Some methods require the Android environment to be set up with specific data or services.
For example, in AmazeFileManager, it is necessary to have an SMB stream server to exercise the methods of class \emph{utils.SmbStreamer}. 
Further, again in AmazeFileManager, to exercise the methods of class GzipExtractor, it is necessary to have files compressed in GZIP format on the filesystem.
\item{\emph{Specific hardware or emulator needed.}} Some methods require installing the App on specific Android hardware or emulators. This is also the case for VLC, where the methods of classes belonging to package \emph{org.videolan.vlc.gui.tv}, can be exercised when the App executes on an Android TV.
Further, the methods of class \emph{TvChannelsKt} can be executed only if a TV tuner (i.e., an hardware device to communicate with a TV provider) is connected to the Android device.
\item{\emph{Specific App settings needed.}} Certain methods can be executed only with specific App settings, but enabling such settings requires a long sequence of actions that neither CALM nor competing approaches can identify. This is the case for VLC, where the methods of classes belonging to package  \emph{org.videolan.vlc.gui.tv} can be exercised without an Android TV, but only after enabling the \emph{TV interface} option in one of the many VLC settings pages. 
\end{itemize}

To discuss complementarity, we further present the differences in target methods covered uniquely by each approach (hereafter, uniquely-covered target methods).
The target methods covered only by one testing approach  account for 3.4\% (i.e., 351 of 10336) of the target methods in our subjects. 
Their distribution is shown in Figure~\ref{fig:target_methods_covered_diff:univocally_covered}. 
\APPR and APE-U are the two approaches yielding the highest number of uniquely-covered target methods, with 29\% (i.e., 101) and 34\% (i.e., 121) of the total, respectively; thus showing some complementarity. 
However, the target methods uniquely covered by APE-U are mainly from two versions of VLC, which account for 71\% and 17\% (i.e., 86 and 26) of the 121 target methods uniquely covered by APE-U, respectively. 
Further, \APPR yields a number of uniquely-covered target methods higher than APE-U 
(i.e., 17 out of 52 subjects for \APPR and only 7 out of 52 for APE-U). 
The effectiveness of \CALM is likely due to the fact that, thanks to App model reuse, \APPR is capable of reaching target Windows and target Widgets that require complex input sequences to be reached (e.g., interacting with widgets in the middle of a list).
Indeed, once \CALM discovers how to reach a target Widget in an App version, it remembers how to do so in future versions.
For example, for Amaze File Manager, \APPR is the only approach reaching the AboutActivity Window and Color Preference Window, and then exercising the target methods in these Windows. Similarly, \APPR is the only approach to exercise target methods in PreferencesCasting and PreferencesSubtitles Windows, in VLC. 

Furthermore, \APPR is the approach covering the largest proportion of target methods across subjects, 70.16\% (i.e., 7252 out of 10336). The other approaches covered a proportion of target methods between 45\% and 66\%. Figure~\ref{fig:target_methods_covered_diff:not_covered_by_each} shows the number of target methods not covered by each approach. 
Figure~\ref{fig:target_methods_covered_diff:not_covered_by_CALM_but_another_and_inverse}, instead, depicts the complementarity between \CALM and each other approach.
Among the target methods missed by \APPR, only 15.5\% of them are covered by other approaches; indeed, orange bars in Figure~\ref{fig:target_methods_covered_diff:not_covered_by_CALM_but_another_and_inverse} are small, which suggests that no approach effectively complements \CALM.
Such proportion is much higher for other approaches (i.e., from 25\% to 54\%), which means that \APPR is  effective in covering target methods missed by other techniques; indeed, blue bars in Figure~\ref{fig:target_methods_covered_diff:not_covered_by_CALM_but_another_and_inverse} are much taller than orange ones. 

Finally, we report that the 468 target methods not covered by \APPR but covered by other approaches are triggered by input types currently not supported by \APPR. 
First, \APPR does not interact with tiny widgets since \APPR assumes these widgets are inactive. This is the case of target methods related to FastScroller in Amaze File Manager, which is a thin stick at the rightmost part of the screen. Second, \APPR does not interact with the popup video player, which is typically the case in VLC, so it could not exercise methods of PopupManager and PopupLayout. 
Third, \APPR is ineffective when a widget supports rich interactions. For example, in the VLC App, swiping on the left side of the playing video will increase or decrease the screen's brightness, depending on the swipe direction, while executing these actions on the right side will change the volume. Some approaches generate (i.e., Monkey) or partially generate (e.g., APE, Fastbot2) pure random events, thus having a chance of triggering these features. \APPR, instead, always swipe widgets in the middle and is thus not able to exercise certain features (e.g., change brightness).
Finally, \APPR does not generate certain system events. For example, it does not turn wi-fi on or off, thus preventing the execution of the methods \emph{onGoOffline()} and \emph{onGoOnline()} in Wikipedia.
Concluding, the limitations above show that technical improvements could further increase \APPR's effectiveness. 

\subsubsection{Limited fault detection capabilities}

\begin{figure}[tb]
    \centering
    \includegraphics[width=\linewidth]{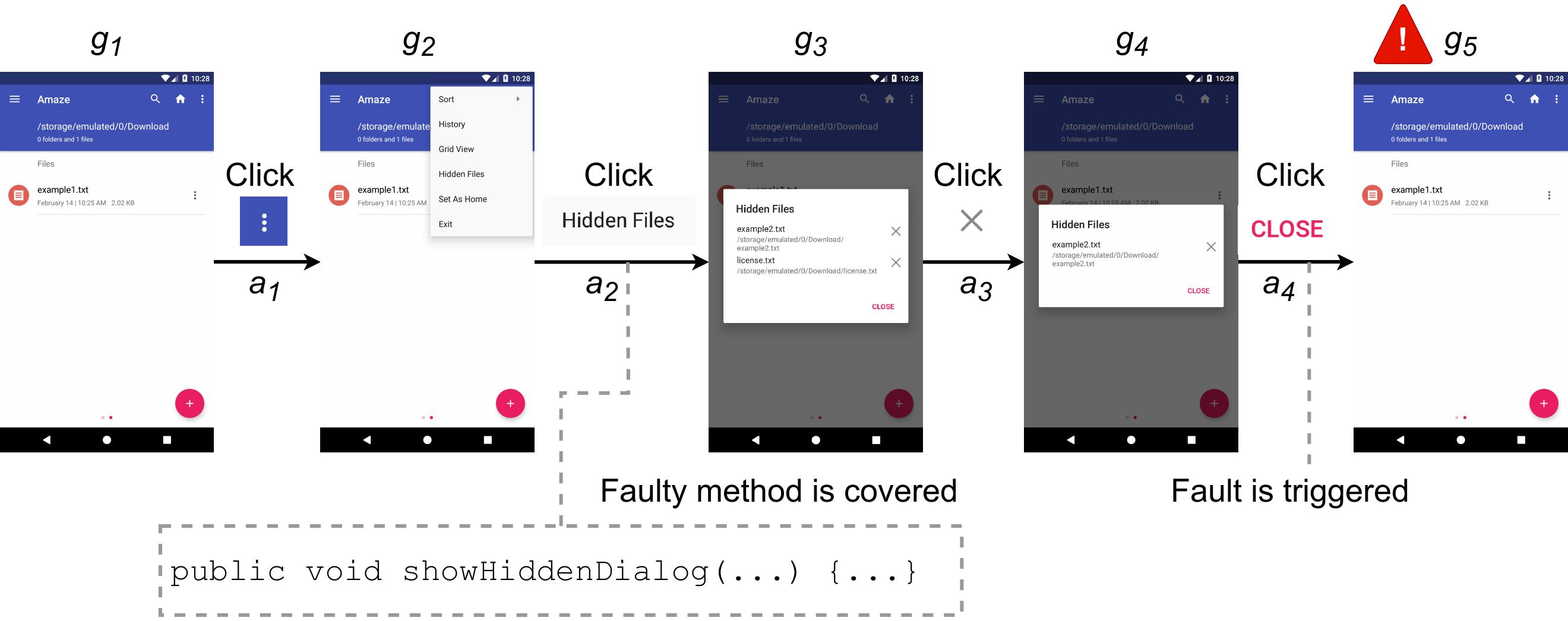}
    \caption{An example of scenario where fault AM07 is detected.}
    \label{fig:am07}
\end{figure}

We identified two reasons preventing fault detection with \CALM and APE-U. First, certain faults can be detected only by inspecting additional output screens besides the one of the action exercising the faulty method. This happens when the faulty method generates data that affects the behaviour of a Window that is different from the one visualized after the execution of the faulty method. 
Such limitation was observed in the case of AM04, AM06, and AM07. 
For example, fault AM07 originates from the missing implementation of an event handler for the ``Close'' button of the ``Hidden Files'' dialog, in the method that creates the dialog (i.e., \emph{showHiddentDialog()}).
Figure~\ref{fig:am07} illustrates a failing execution. The faulty method (i.e., the method showHiddenDialog()) is always exercised when the ``Hidden Files'' dialog is opened (i.e., action $a_2$). However, to trigger fault AM07, it is necessary to remove at least one hidden file from the list in the ``Hidden Files'' dialog (i.e.,  action $a_3$), and then close the dialog (i.e., action $a_4$). A failure can then be observed in $g_5$ because the removed hidden file is not present in the file list. However, the unique target action is $a_2$, which covers the faulty method but inspecting the screenshots taken before and after $a_2$ (i.e., $g_2$ and $g_3$) is insufficient to detect the fault. Instead, engineers should inspect the screenshot taken before and after $a_4$ (i.e., $g_4$ and $g_5$).
To address such limitation, it might be useful to report to the end-user not only the output of a unique target action, but also related outputs, which would significantly increase testing cost. However, based on our results for RQ4, we noted that such limitation is observed with actions that change what is displayed in their parent Window. 
Therefore, to increase fault detection capabilities, we may extend \CALM to display not only the screenshot taken after the target action but also the screenshot with the next visualized Window.
Note that even if we double the number of outputs to be inspected with \CALM (i.e., it always reports the screenshot produced after the target action and the screenshot with the next Window visualized after the target action, for every target action), it remains the least expensive approach, with an average of 88 output screens to be inspected versus 878 of the cheapest competitor, \ATUA.
 
The second reason why faults are not detected by \CALM and APE-U is that 
exercising the faulty code is not sufficient to trigger the failure when it requires specific inputs or the App to be in a specific state. Specific inputs are needed when the fault consists of missing instructions (e.g., 
AM02 is due to a missing function invocation) or when the fault consists of wrong parameters passed to a delegate method returning the wrong result (e.g., AM05 happens because the string ".txt" is passed to a filter function). A specific App state is required when the fault consists of a violation of the preconditions for the execution of a specific portion of code; this is what happens for AM02, where the ``Deselect All'' menu item correctly exercises the code that deselects all the items in a list, while the ``Select All'' menu item erroneously executes the same code if all the items are already selected. Please note that, in all these cases, \CALM triggers the failure but the output screenshot with the failure is not selected for inspection, which indicates that, to detect faults, it is not sufficient to report only the screenshot captured the first time a target action is exercised (i.e., what we call UTA). Instead, some contextual information should be considered to determine if a target action should be considered for inspection; based on the cases above, the nature of the inputs and the current abstract state should be taken into account. For example, we could detect AM05 by distinguishing between cases in which the target action is triggered with inputs containing a string appearing in the source code (i.e., ".txt") from other cases. We could detect AM02 by reporting all the output screens obtained by executing the target action from distinct AbstractStates. 

The investigation of all the strategies suggested above goes beyond the scope of this paper, which is about model reuse effectiveness, not output selection. Further, despite these limitations, \CALM could detect half of the faults with a test budget of two hours (i.e., when \CALM is executed twice, see \ref{sec:rq4:results}).

\MAJOREND{}

\subsection{Threats to validity}

\emph{Internal validity.}
To minimize \emph{implementation errors}, we have carefully tested \APPR before running our experiments. For the selected competing state-of-the-art tools, we relied on the versions released by their authors, which had been extensively used in related work. 

\emph{Conclusion validity.}
{To avoid violating the \emph{assumptions of parametric statistical tests}, we rely on a non-parametric test and effect size measure (i.e., Mann Whitney U-test and the Vargha and Delaney’s $A_{12}$ statistics, respectively).}
{To ensure \emph{reliability}, our measurements (i.e., code coverage) have been collected through widely used, open-source tools.}

\emph{Construct validity.}
The constructs considered in our work are effectiveness and cost. 
Effectiveness is measured through two reflective indicators, which are target method coverage and target instruction coverage. We rely on code coverage because it is a common measure of effectiveness for functional testing~\cite{Choudhary-AutomatedTestInputGeneration-ASE-2015,DiMartino2020}.
Cost is measured in terms of the number of target actions whose effects (i.e., resulting App screens) should be inspected to determine test outcome, as discussed in Section~\ref{sec:rq1:design}.

\emph{External validity.}
We have considered seven popular Apps, used in related work and in the empirical assessment of \ATUA, which enabled fair comparison to discuss the coverage improvements enabled by \CALM. For each App, we considered up to ten App versions, based on their availability, for a total of 52 App versions tested. 
The considered Apps are diverse in terms of features, the overall number of instructions, and updated instructions between versions. 

To account for randomness, we tested each App version ten times with every testing tool considered. Despite the high computational cost (6940 test execution hours, in total), this enabled us to derive solid statistical results for the comparison of  different tools.

In our experiments we considered only Apps for the Android operating system, which is the platform most targeted by research work.
The choice of relying on Android Apps enabled the comparison of \APPR with six competing tools, all of which work only with Android Apps.
However, our approach does not rely on any assumption restricting its applicability to Android. To drive testing, it requires code coverage, which is measurable on any platform, and GUI Trees. CALM extracts GUI Trees by relying on the Android UIAutomator API. Similar features are provided by Appium~\cite{appium}, which works with both iOS and Windows OS. Also, Harmony OS may provide a UIAutomator-like API. 

\begin{figure*}
    \begin{subfigure}[b]{7cm}
        \includegraphics[width=7cm]{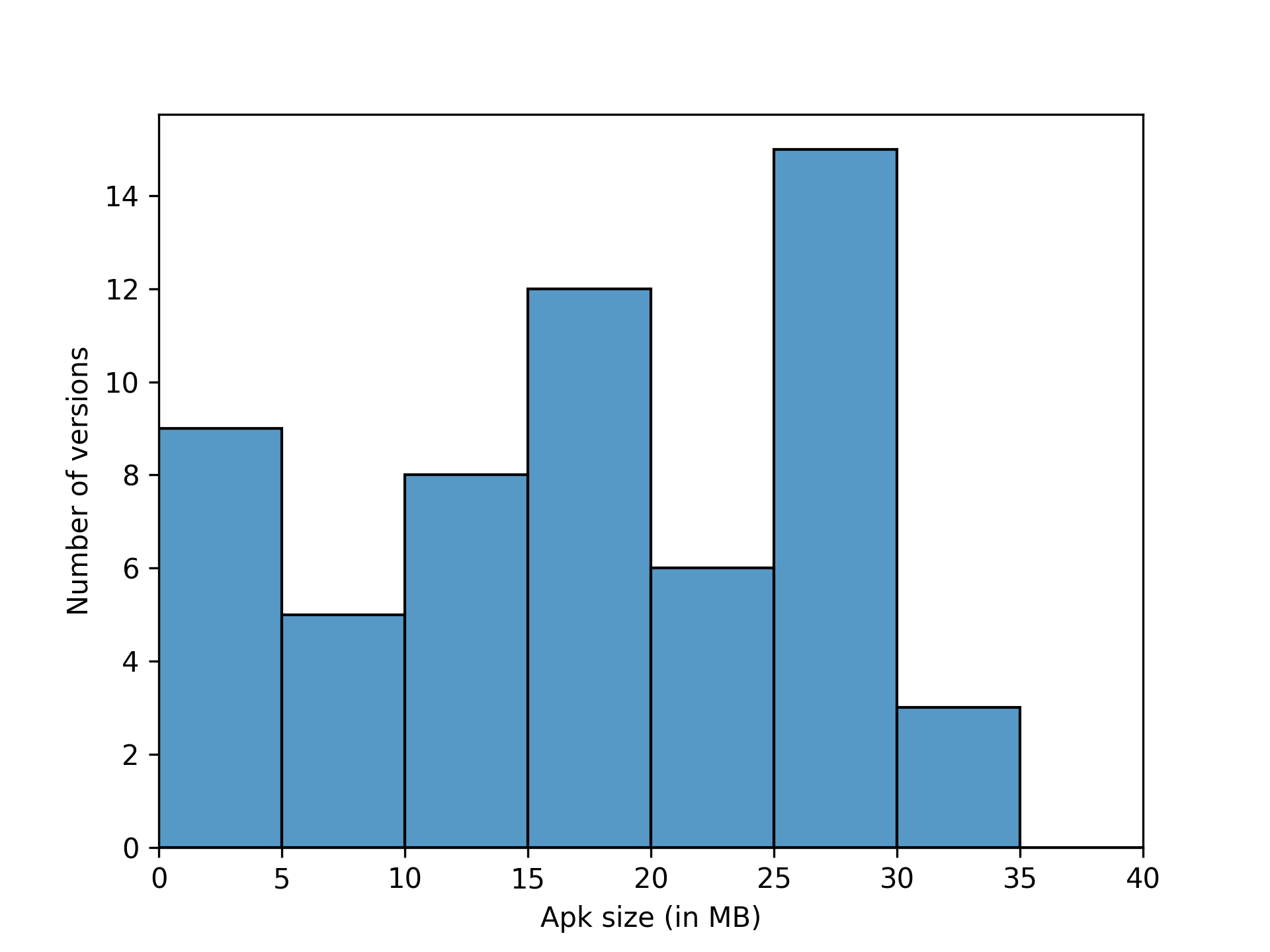}
        \caption{APK disk occupation}
        \label{fig:APK:disk:occupation:subjects}
    \end{subfigure}\begin{subfigure}[b]{7cm}
        \centering
        \includegraphics[width=7cm]{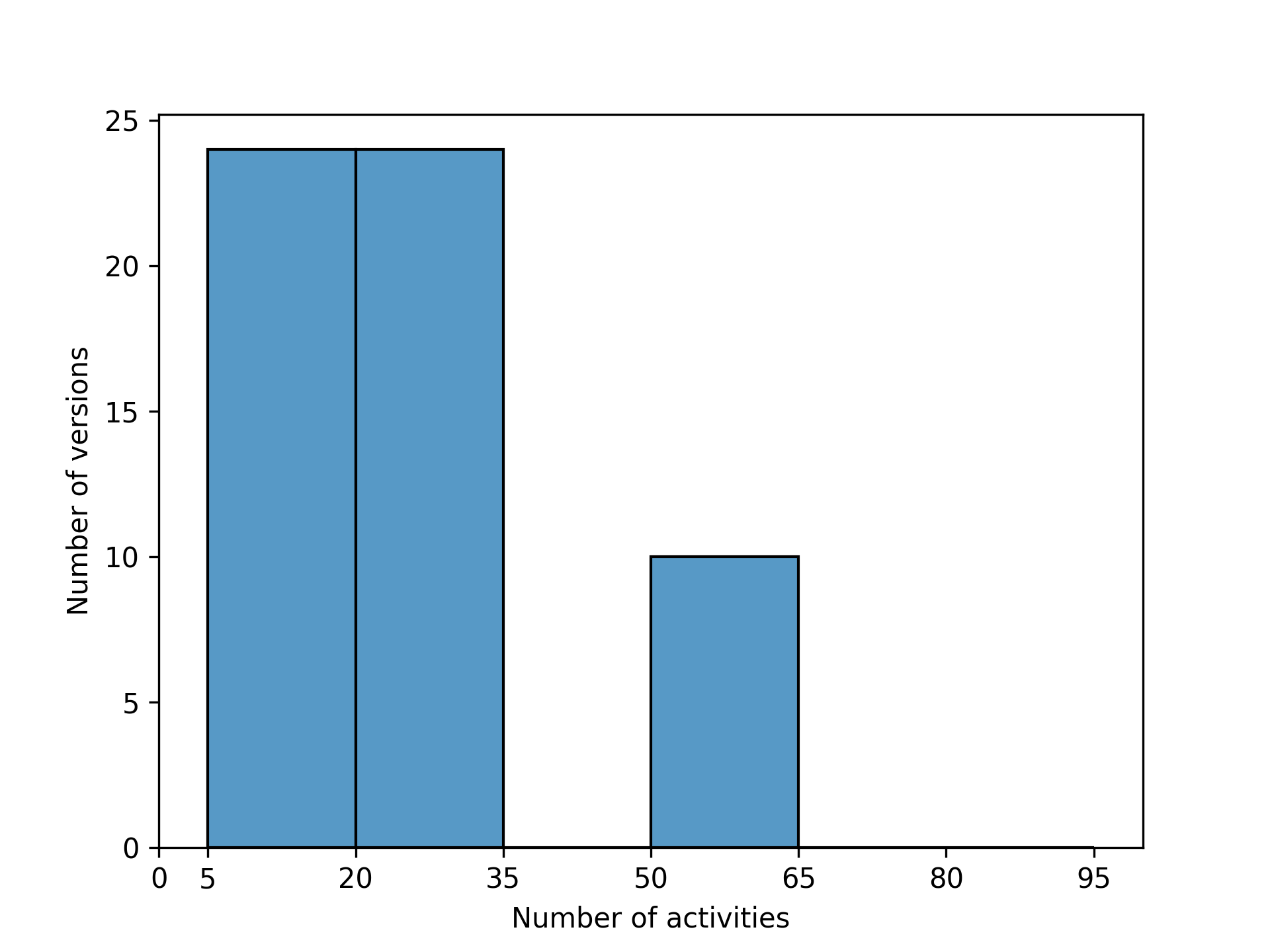}     
        \caption{Activities}
        \label{fig:APK:disk:occupation:subjects}
    \end{subfigure}
       \caption{Distribution of APK disk occupation (Mb) and activities for our subject Apps.}
        \label{fig:size:subjects}
\end{figure*}

\begin{figure*}
    \begin{subfigure}[b]{7cm}
        \includegraphics[width=7cm]{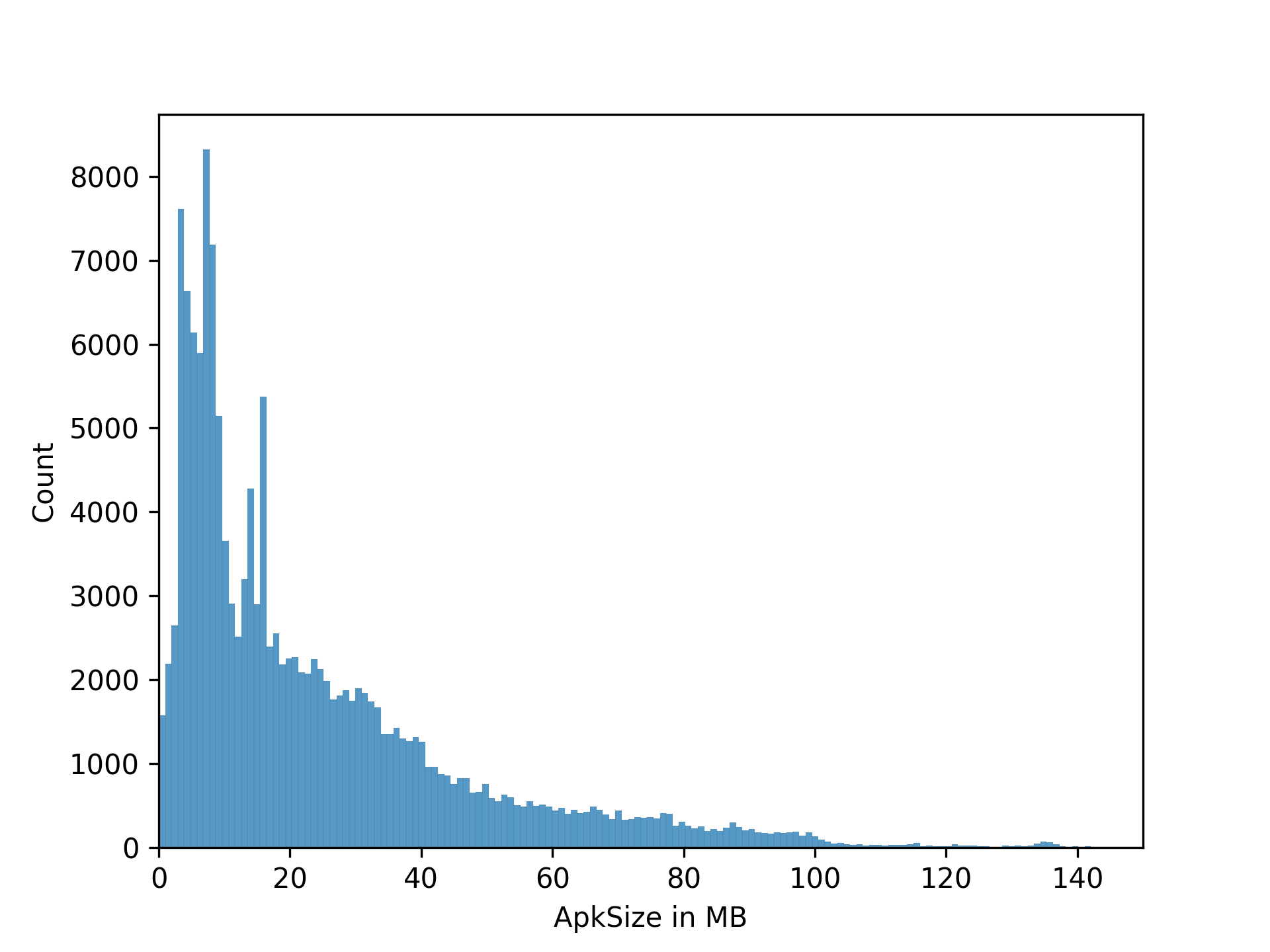}
        \caption{2020}
        \label{fig:APK:disk:occupation:2020}
    \end{subfigure}\begin{subfigure}[b]{7cm}
        \centering
        \includegraphics[width=7cm]{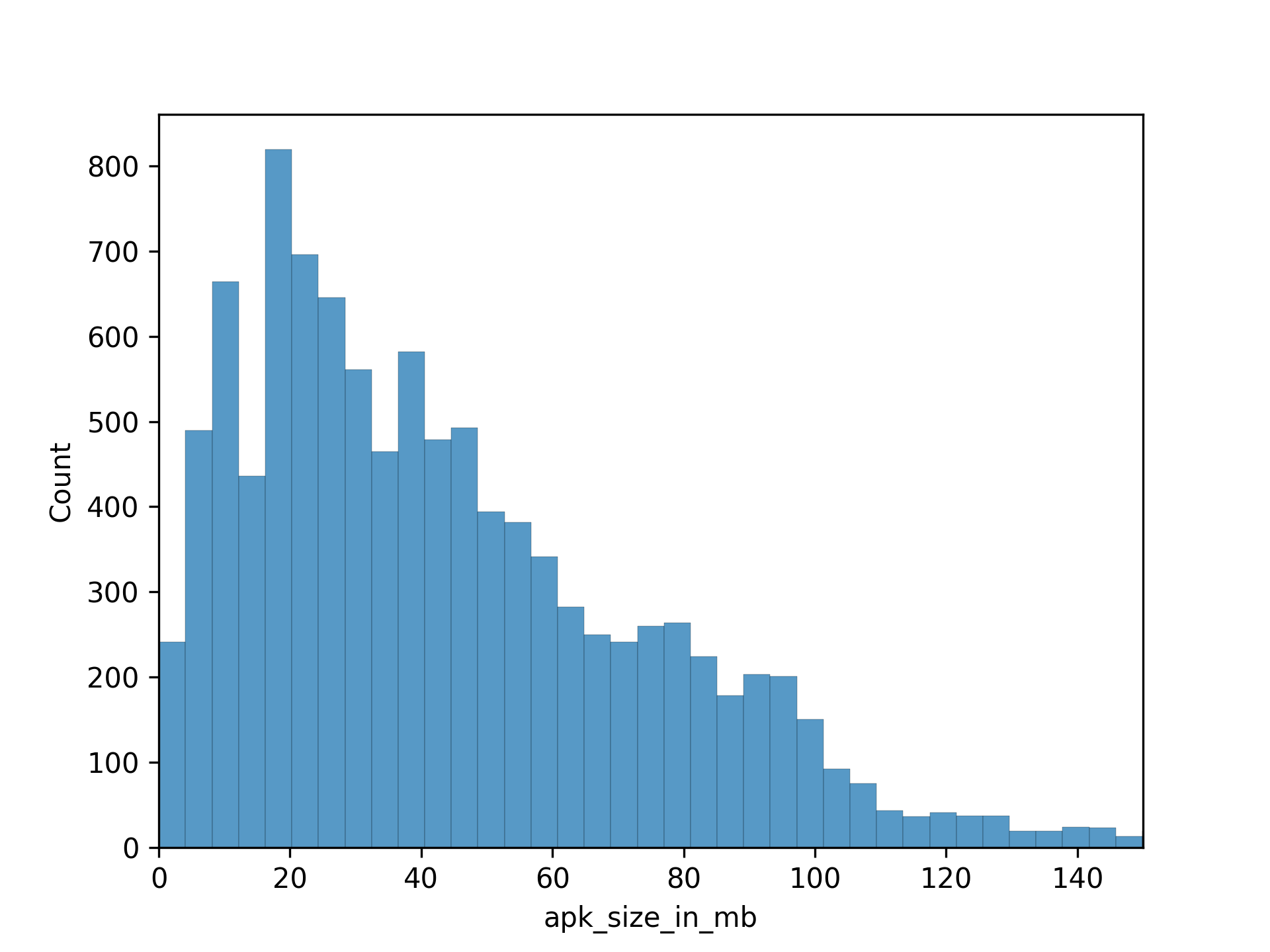}     
        \caption{2023}
        \label{fig:APK:disk:occupation:2023}
    \end{subfigure}
       \caption{Distribution of APK disk occupation (Mb) for Android Apps in 2020 and 2023.}
        \label{fig:APK:disk:occupation}
\end{figure*}

\MAJOR{R1.3}{The size of the tested Apps may affect the outcome of our approach, e.g., large Apps may not be successfully processed by Gator. Since assessing the scalability of Gator is out of the scope of this paper, to address treats to  generalizability, we aim to demonstrate that the selected Apps have a typical size for Apps published on Android markets. Given that App size can be captured in terms of APK disk occupation and number of activities, we aim to demonstrate that the distribution of these two metrics' values for our subject Apps match the distribution observed with all the Apps on Android markets. 
Please note that our subject Apps belong to the set of Apps selected for the empirical assessment of ATUA and was released in 2020.
Figures~\ref{fig:size:subjects}~(a) and ~\ref{fig:size:subjects}~(b) provide plots with the distribution of APK disk occupation (Mb) and number of activities for our subject Apps. Disk occupation ranges from 2.4 Mb (Activity Diary version 105) to 32.3 Mb (Yahooweather version 1.20.7), the number of activities ranges from 7 to 64.
 Figure~\ref{fig:APK:disk:occupation} shows the distribution of APK disk occupation of all the Apps released on the main Android markets in 2020 and 2023; we derived the boxplots in Figure~\ref{fig:APK:disk:occupation} from the data reported by AndroZoo~\cite{Allix:2016:ACM:2901739.2903508}, which is the largest repository of Android Apps.}

\MAJORBEGIN{}
Based on our analysis, in 2020, 
107,000 Apps (71\%) had an APK disk occupation between 2 Mb and 32 Mb, thus showing that the Apps selected for \ATUA were representative of most Apps in the Android markets in 2020. In 2023, AndroZoo collected 10,600 Apps, with 4,400 (41.5\%) Apps being between 2 Mb and 32 Mb in size. Though the median App size increased from 18 Mb in 2020 to 38 Mb in 2023, this data shows that our subject Apps are still representative of the size of almost half of the Apps on the Android markets, thus supporting the generalizability of our empirical results.

\begin{figure*}
        \includegraphics[width=7cm]{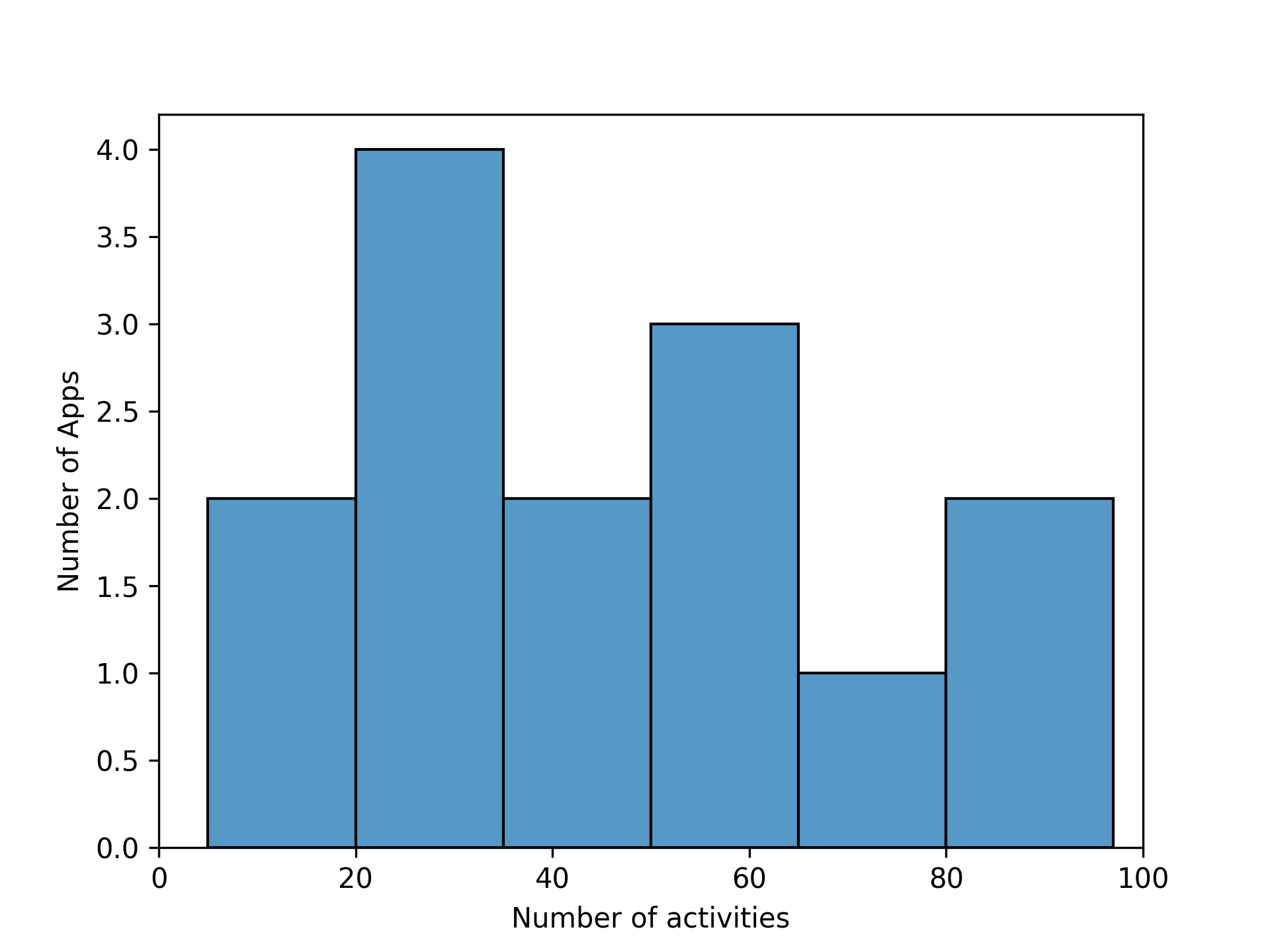}
       \caption{Distribution of activities across Apps tested in related work~\cite{Gu:APE:ICSE:2019}}
            \label{fig:res:activities:APE}
\end{figure*}

Since AndroZoo does not collect information about Apps' activities, we report in Figure~\ref{fig:res:activities:APE} the distribution of the number of activities for the Apps considered in the empirical evaluation of APE~\cite{Gu:APE:ICSE:2019}, which is the only paper among the ones selected as our comparison baseline reporting such data for each subject App. Figure~\ref{fig:res:activities:APE} shows that the number of activities for our subject Apps is similar to that of APE, except for three Apps in the APE set that have more than 64 activities. Such result, again, demonstrates that our findings likely generalize to a broad set of Apps.
\MAJOREND{}


\section{Related Work}
\label{sec:related}

App testing tools differ with respect to their input selection strategy~\cite{Linares:ICSME:2017,Amalfitano:AndroidTestingSurvey:SQJ:2018}; most rely on random\cite{monkey}, model-based \cite{Borges-Droidmate2-ASE-2018, stoat17,Gu:APE:ICSE:2019,combodroid}, search-based~\cite{sapienz}, deep~\cite{Li:Humanoid:2019,DeepGUI,MuBot} and reinforcement learning~\cite{Koroglu:QBE:2018,Pan:QTesting:2020,Romdhana:ARES:2022,Fastbot2} strategies.

 Model-based approaches dynamically construct a model for the App under test, which is used to drive testing. DM2~\cite{Borges-Droidmate2-ASE-2018} and APE~\cite{Gu:APE:ICSE:2019}, similar to ATUA, construct the model while verifying the App. 
 Stoat~\cite{stoat17} and ComboDroid~\cite{combodroid}, instead, first  construct an App model, which is then used to perform testing.
 Although the models derived by such approaches may be reusable across versions, different from \CALM, none of the approaches above rely on model reuse. We demonstrated that \APPR's heuristics enable effective reuse of ATUA's App models; future work includes assessing \APPR's heuristics when integrated into other SOTA tools. 
 
 Although other heuristics to improve App testing exist, ours are the first being tailored for update testing. For example, ProMal (published in 2022) improves the precision of WTGs by relying on dynamic analysis, which is a strategy integrated into ATUA (2021). 
 Similar to CALM's layout-guarded AbstractTransitions, some approaches extend WTGs with backward transitions capturing when an input brings the App to a previous Window~\cite{yang-jase18,Zhang:2018:Launch}; however, they model Window transitions, not AbstractTransitions, which are instead used by CALM to maximize testing effectiveness. 
 

The main output of random and search-based approaches are test cases, which, unlike App models, cannot be reused to test new features (e.g., to reach required Window states). That is why we extended a model-based approach (i.e., \ATUA).

As for reinforcement learning approaches, only Fastbot2~\cite{Fastbot2} reuses models across versions. It leverages a probabilistic model capturing the likelihood that each Action triggerable in a Window reaches another Window. Our results show that it is inappropriate for update testing; further, its source code is not available. Similarly, deep learning (DL) approaches do not target updated code and, in our experiments, \CALM outperformed the most cited and available DL-based tool (i.e., Humanoid).

Other techniques, different from \CALM, update GUI test scripts but do not automatically test Apps.  
ATOM~\cite{Li:Atom:2017} and CHATEM~\cite{Chang:CHATEM:2018} rely on a base App model and a \emph{Difference} model constructed either manually (ATOM) or semi-automatically (CHATEM). GUIDER~\cite{Xu:GUIDER:2021}, instead, does not require a difference model but, 
while executing test scripts, compares the App screens of the base and updated App version to identify actions leading to the expected states. 

Xiong et al. have recently conducted an empirical study of 399 functional faults in Android Apps~\cite{FuntionalBugsAndroid}. Their results highlight that most of the functional faults require visual inspection to be detected. Indeed, they report that only 30\% of the faults lead to crashes. Of the remaining, only 3\% are related to energy consumption, the rest is probably detectable only through visual inspection; indeed, content related issues account for 21\%, structure related issues 40\%, incorrect interaction 19\%, functionality not taking effect 12\%, and unresponsive UI element 5\%.
Further, Xiong et al. also report that feature agnostic oracles (e.g., looking for overlapping UI elements, data loss, and App freezing) discover 30\% of non-crashing functional failures; however, in their experiments, existing tools implementing feature agnostic oracles (i.e., Genie~\cite{Genie}, Odin~\cite{ODIN}, IFixDataLoss~\cite{IFixDataLoss}, ITDroid~\cite{ITDroid}, and SetDroid~\cite{SetDroid}) could, overall, detect only 6\% of such faults. In other words, 84\% of non-crashing failures can be detected only through visual inspection, which further motivates our work. Finally, Xiong et al. propose RegDroid, which implements a form of differential testing to detect regression faults (i.e., it executes a random sequence of actions on two App versions and verifies if the output screens include a same randomly selected interactable widget); unfortunately, their results show a false positive rate above 60\%. Since regression faults are a subset of all the possible functional faults in an App, engineers still need to inspect App outputs for non-regression faults; therefore, for engineers, it would likely be more effective to simply inspect all the outputs generated by an App.
Consequently, an approach like \CALM, which targets modified functionalities, thus leading to a limited number of outputs to inspect, may be practically useful. Studying how the inputs generated by \CALM might be used for differential testing is part of our future work.

\section{Conclusion}
\label{sec:conclusion}

We presented \APPR, a technique to efficiently test App updates by relying on models learned with previous App versions. 
It relies on static analysis to identify GUI components modified across versions and adapt App models accordingly (e.g., reuse abstract states for renamed Windows); further, it integrates four heuristics addressing the limitations of model inference that are exacerbated in the presence of model reuse: It infers \emph{layout-guarded abstract transitions}, which deal with non-deterministic transitions; it derives \emph{probabilistic Action sequences}
to deal with \emph{state explosion}; it detects model states that are new but compatible with previously executed Action sequences (i.e., \emph{backward-equivalent}); it relies on \emph{online and offline model refinement} to identify and remove obsolete states.

Our empirical evaluation shows that \APPR leads to a coverage of updated methods and instructions that is higher than the second best SOTA approach by 6 percentage points (pp) for a one-hour test budget. That difference keeps steadily widening as the test budget increases and is larger for smallest updates (13 pp), which are the most frequent. 




\begin{acks}
This project has received funding from Huawei Technologies Co., Ltd, China, and 
 by the NSERC Discovery and Canada Research Chair programs.
Experiments presented in this paper were carried out using the Grid'5000 testbed, supported by a scientific interest group hosted by Inria and including CNRS, RENATER and several Universities as well as other organizations (see https://www.grid5000.fr).
\end{acks}


\bibliographystyle{ACM-Reference-Format}
\bibliography{CALM}






\end{document}